\documentclass[11pt,english]{article}
\usepackage[T1]{fontenc}
\usepackage[latin1]{inputenc}
\usepackage{babel}

\makeatletter

\providecommand{\LyX}{L\kern-.1667em\lower.25em\hbox{Y}\kern-.125emX\@}

\usepackage{amsfonts}

\makeatother
\begin{document}
\begin{flushright}

MPP-2003-79

\end{flushright}

{\centering \textbf{\LARGE Construction of Gauge Theories on Curved
Noncommutative Spacetime}\LARGE \par}

\textbf{\bigskip}

\bigskip

{\centering Wolfgang Behr%
\footnote{behr@theorie.physik.uni-muenchen.de
}, Andreas Sykora%
\footnote{andreas@theorie.physik.uni-muenchen.de
}\par}

\bigskip

{\centering Max-Planck-Institut f\"ur Physik, \par}

{\centering F\"ohringer Ring 6, 80805 M\"unchen, Germany\par}

\begin{abstract}
We present a method where derivations of \( \star  \)-product algebras are
used to build covariant derivatives for noncommutative gauge theory. We write down a noncommutative action
by linking these derivations to a frame field induced by a nonconstant
metric. An example is given where the action reduces in the classical
limit to scalar electrodynamics on a curved background.
We further use the Seiberg-Witten map to extend the formalism to arbitrary gauge groups.
A proof of the existence of the Seiberg-Witten-map for an abelian
gauge potential is given for the formality \( \star  \)-product.
We also give explicit formulas for the Weyl ordered \( \star  \)-product and its Seiberg-Witten-maps
up to second order. 
\end{abstract}


\newpage

\tableofcontents{}

\newpage

\section{Introduction}

One hope associated with the application of noncommutative geometry
in physics is a better description of quantized gravity. At least
it should be possible to construct effective actions where traces
of this unknown theory remain. If one believes that quantum gravity
is in a sense a quantum field theory, then its observables are operators
on a Hilbert space and therefore elements of an algebra. Some properties
of this algebra should be reflected in the noncommutative geometry
the effective actions are constructed on. As the noncommutativity
should be induced by background gravitational fields, the classical
limit of the effective actions should reduce to actions on curved
spacetimes \cite{Madore:1997ta,Connes:1996gi}.

In this paper we will investigate noncommutative geometry formulated
in the \( \star  \)-product formalism, where gauge theory can be constructed in a particularly convenient way on noncommutative spacetime. 

The case of an algebra with constant commutator has been extensively
studied. This theory reduces in the classical limit to a theory on
a flat spacetime. Therefore it is necessary to develop concepts working
with more general algebras\footnote{As an example for the treatment of a special algebra, see also the recent paper \cite{Meyer:2003wj}, where gauge theory on the $E_q(2)$-covariant plane is studied.}, since one would expect that curved backgrounds
are related to algebras with nonconstant commutation relations. We
present here a method using derivations of \( \star  \)-product algebras
to build covariant derivatives for noncommutative gauge theory. We are able to write down a noncommutative action by linking these derivations
to a frame field induced by a nonconstant metric. An example is given
where the action reduces in the classical limit to scalar electrodynamics
on a curved background.

Nonexpanded theories can
only deal with \( U(n) \)-gauge groups, but using Seiberg-Witten-maps
relating noncommutative quantities with their commutative counterparts
makes it possible to consider arbitrary nonabelian gauge groups \cite{Seiberg:1999vs,Madore:2000en,Jurco:2001rq}. We therefore extend our formalism to arbitrary gauge groups by introducing Seiberg-Witten maps. 
A proof of the existence of the Seiberg-Witten-map for an abelian
gauge potential is given for the formality \( \star  \)-product.
We also give explicit formulas for the Weyl ordered \( \star  \)-product and its Seiberg-Witten-maps
up to second order.

\section{The general formalism}

\subsection{Classical gauge theory}

First let us recall some properties of a general classical gauge theory.
A non-abelian gauge theory is based on a Lie group with Lie algebra\begin{equation}
[T^{a},T^{b}]=i\, f^{ab}{}_{c}T^{c}.\end{equation}
Matter fields transform under a Lie algebra valued infinitesimal parameter\begin{equation}
\label{com_g_field}
\delta _{\alpha }\psi =i\alpha \psi ,\, \, \, \, \, \, \, \, \, \alpha =\alpha _{a}T^{a}
\end{equation}
in the fundamental representation. It follows that\begin{equation}
\label{com_consitency_condition}
(\delta _{\alpha }\delta _{\beta }-\delta _{\beta }\delta _{\alpha })\psi =\delta _{-i[\alpha ,\beta ]}\psi .
\end{equation}
The commutator of two consecutive infinitesimal gauge transformation
closes into an infinitesimal gauge transformation. Further a Lie algebra
valued gauge potential is introduced with the transformation property\begin{eqnarray}
a_{i} & = & a_{ia}T^{a},\nonumber \\
\delta _{\alpha }a_{i} & = & \partial _{i}\alpha +i[\alpha ,a_{i}].\label{com_gt_a} 
\end{eqnarray}
With this the covariant derivative of a field is\begin{equation}
D_{i}\psi =\partial _{i}\psi -ia_{i}\psi .\end{equation}
The field strength of the gauge potential is defined to be the commutator
of two covariant derivatives\begin{equation}
iF_{ij}=[D_{i},D_{j}]=\partial _{i}a_{j}-\partial _{j}a_{i}-i[a_{i},a_{j}].\end{equation}
The last equations can all be stated in the language of forms. For
this a connection one form is introduced\begin{equation}
a=a_{ia}T^{a}dx^{i}.\end{equation}
The covariant derivative now acts as\begin{equation}
D\psi =d\psi -ia\psi .\end{equation}
The field strength becomes a two form\begin{equation}
F=da-ia\wedge a.\end{equation}

\subsection{Commutative actions with the frame formalism}

In this section we want to recall some aspects of classical differential
geometry. Suppose we are working on a \( n \)-dimensional manifold
\( M \). Then there are locally \( n \) derivations \( \partial _{\mu } \)
which form a basis of the tangent space \( TM \) of the manifold.
The derivations all fulfill the Leibniz rule on two functions. If
we make a local basis transformation on \( TM \) then this {}``frame''
can always be written locally as \begin{equation}
e_{a}=e_{a}{}^{\mu }(x)\partial _{\mu },\end{equation}
where \( e_{a}{}^{\mu }(x) \) has to be invertible (\( e_{\nu }{}^{a}e_{a}{}^{\mu }=\delta _{\nu }^{\mu } \)).
Since forms are dual to vector fields, they may be evaluated on a
frame. For the covariant derivate we get\begin{equation}
(D\psi )(e_{a})=e_{a}\psi -ia_{a}\psi \end{equation}
where\begin{equation}
a_{a}=a(e_{a}).\end{equation}
The field strength becomes\begin{equation}
f(e_{a},e_{b})=f_{ab}=e_{a}a_{b}-e_{b}a_{a}-a([e_{a},e_{b}])-i[a_{a},a_{b}].\end{equation}
It is well known that in Riemannian geometry it is always possible
to find a frame where the metric is constant\begin{equation}
\eta _{ab}=e_{a}{}^{\mu }e_{b}{}^{\nu }g_{\mu \nu }.\end{equation}
Since in scalar electrodynamics we do not need a spin connection,
it is simple to write down its action on an curved manifold with the
frame formalism\begin{equation}
\mathcal{S}=\int d^{n}x\, e\, (-\frac{1}{4}\eta ^{ab}\eta ^{cd}f_{ac}f_{bd}+\eta ^{ab}D_{a}\overline{\phi }D_{b}\phi +m^{2}\overline{\phi }\phi ).\end{equation}
 Here \begin{equation}
e=(\det \, e_{a}{}^{\mu })^{-1}=\sqrt{\det \, (g_{\mu \nu })}\end{equation}
is the measure function for the curved manifold.

\subsection{Noncommutative gauge theory}

In our approach to studying physics in the noncommutative realm, one deforms the commutative
algebra of functions on a space to a noncommutative one. This noncommutative
algebra of functions we call noncommutative space, noncommutative
objects are written with a hat. We want this deformation to be controlled
by some parameter so that in some limit we can get back a commutative
space. The same we expect from theories built on a noncommutative
space: In the commutative limit they should reduce to a meaningful
commutative theory. For a noncommutative space where the commutator
of the coordinates is a constant, the commutative limit is the usual
gauge theory on flat spacetime. But as the noncommutativity should
be related to gravity, gauge theory on a curved spacetime seems to
be a more natural limit for theories on noncommutative spaces with
more complicated, non-constant commutators.

In a gauge theory on a noncommutative space, fields should again
transform like (\ref{com_g_field})\begin{equation}
\label{nc_gt_field}
\hat{\delta }_{\hat{\Lambda }}\hat{\Psi }=i\hat{\Lambda }\hat{\Psi }.
\end{equation}
Again the commutator of two gauge transformations should be a gauge transformation\begin{equation}
\label{nc_consitency_condition}
(\hat{\delta }_{\hat{\Lambda }}\hat{\delta }_{\hat{\Gamma }}-\hat{\delta }_{\hat{\Gamma }}\hat{\delta }_{\hat{\Lambda }})\hat{\Psi }=\hat{\delta }_{-i[\hat{\Lambda },\hat{\Gamma }]}\hat{\Psi }.
\end{equation}

This is only the case for \(U(n)\) gauge groups, but general gauge groups can be implemented by using Seiberg-Witten maps (see chapter \ref{swmap}).
Since multiplication of a function with a field is not again a covariant
operation, we are forced to introduce a covariantizer with the transformation
property \begin{equation}
\label{nc_gt_D}
\hat{\delta }_{\hat{\Lambda }}D(\hat{f})=i[\hat{\Lambda },D(\hat{f})].
\end{equation}
From this it follows that\begin{equation}
\label{nc_gt_field}
\hat{\delta }_{\hat{\Lambda }}(D(\hat{f})\hat{\Psi })=i\hat{\Lambda }D(\hat{f})\hat{\Psi }.
\end{equation}
If we covariantize the coordinate functions \( \hat{x}^{i} \) we
get covariant coordinates \begin{equation}
\hat{X}^{i}=D(\hat{x}^{i})=\hat{x}^{i}+\hat{A},^{i}\end{equation}
where the gauge field now transforms according to \begin{equation}
\label{nc_gt_covcord}
\hat{\delta }_{\hat{\Lambda }}\hat{A}^{i}=-i[\hat{x}^{i},\hat{\Lambda }]+i[\hat{\Lambda },\hat{A}^{i}].
\end{equation}

Unluckily, this does not have a meaningful commutative limit, a problem
that can only be fixed for the canonical case (i.e. \( [\widehat{x}^{i},\widehat{x}^{j}]=i\theta ^{ij} \)
with \( \theta  \) a constant) and invertible \( \theta  \). 

For noncommutative algebras where we already have derivatives with
a commutative limit, it therefore seems natural to gauge these. But
due to their nontrivial coproduct the resulting gauge field would
have to be derivative-valued to match the rather awkward behaviour
under gauge transformations \cite{Dimitrijevic:2003pn}. The physical reason for this might be
the following: The noncommutative derivatives are in general built
to reduce to derivatives on flat spacetime, which might not be the
correct commutative limit.

We therefore advocate a solution using derivations that will later
on (see section \ref{Generalization to NC geometry}) be linked to
derivatives on curved spacetime:

If we have a derivation \( \hat{\partial } \) of the algebra, i.e.
\begin{equation} 
\hat{\partial } (\hat{f}\hat{g})=(\hat{\partial }\hat{f})\hat{g}+ \hat{f} (\hat{\partial }\hat{g}),
\end{equation}
we can introduce a noncommutative gauge parameter \( \hat{A}_{\hat{\partial }} \)
and demand that the covariant derivative (or covariant derivation)
of a field \begin{equation}
\hat{D}\hat{\Psi }=(\hat{\partial }-i\hat{A}_{\hat{\partial }})\hat{\Psi }\end{equation}
again transforms like a field \begin{equation}
\hat{\delta }_{\hat{\Lambda }}\hat{D}\hat{\Psi }=i\hat{\Lambda }\hat{D}\hat{\Psi }.\end{equation}
From this it follows that \( \hat{A}_{\hat{\partial }} \) has to
transform like \begin{equation}
\label{nc_gt_A}
\hat{\delta }_{\hat{\Lambda }}\hat{A}_{\hat{\partial }}=\hat{\partial }\hat{A}_{\hat{\partial }}+i[\hat{\Lambda },\hat{A}_{\hat{\partial }}].
\end{equation}
This is the transformation property we would expect a noncommutative
gauge potential to have. 
If we have an involution on the algebra, we can demand
that the gauge potential is real \( \hat{A}_{\hat{\partial }}=\overline{\hat{A}_{\hat{\partial }}} \)
and the field \( \overline{\hat{\Psi }} \) transforms on the right
hand side. In this case expressions of the form\begin{eqnarray}
\overline{\hat{\Psi }}\hat{\Psi } & \mbox {and} & \overline{\hat{D}\hat{\Psi }}\hat{D}\hat{\Psi }
\end{eqnarray}
become gauge invariant quantities.

\subsection{Derivations and Forms}

We have seen that in order to construct noncommutative gauge theory in our approach, we need derivations on the algebra. As we want to use \(\star\)-products to represent the algebra, we have to investigate the derivations of such a \(\star\)-product algebra.
We will be able to identify derivations of \( \star  \)-products with Poisson
vector fields of the Poisson structure associated with the \( \star  \)-product.
To be more explicit, let us assume that \( X \) is a Poisson vector
field\begin{equation}
X^{i}\partial _{i}\{f,g\}=\{X^{i}\partial _{i}f,g\}+\{f,X^{i}\partial _{i}g\}.\end{equation}
Then there exists a polydifferential operator \( \delta _{X} \)
with the following property
\begin{equation}
\delta _{X}(f\star g)=\delta _{X}f\star g+f\star \delta _{X}g.\end{equation}

Such a map \( \delta  \) from the vectorfields to the differential operators, which maps the derivations of the Poisson manifold \( T_{\pi }M=\{X\in TM|[X,\pi ]_{S}=0\} \) to the derivations of the \( \star  \)-product \( D_{\star }M=\{\delta \in D_{poly}|[\delta ,\star ]_{G}=0\} \), can be constructed both for the Formality \(\star\)-product (see \ref{Formality map}) and the Weyl ordered \(\star\)-product (see \ref{weyl derivations}). Here we want to investigate the general properties of such a map \(\delta\). For this we expand it on a local patch in terms
of partial derivatives\begin{equation}
\delta _{X}=\delta ^{i}_{X}\partial _{i}+\delta ^{ij}_{X}\partial _{i}\partial _{j}+\cdots .\end{equation}
Due to its property to be a derivation, \( \delta _{X} \) is completely determined by the first term \( \delta ^{i}_{X}\partial _{i} \). This means that if the first term is zero, the other terms have to vanish, too. If further \( e \) is an arbitrary derivation of the
\( \star  \)-product, there must exist a vector field \( X_{e} \)
such that\begin{equation}
\delta _{X_{e}}=e.\end{equation}
If \( X,Y\in T_{\pi }M \), then \( [\delta _{X},\delta _{Y}] \)
is again a derivation of the \( \star  \)-product and we can conclude
that \begin{equation}\label{lie_delta}
[\delta _{X},\delta _{Y}]=\delta _{[X,Y]_{\star }},\end{equation}
where \( [X,Y]_{\star } \) is a deformation of the ordinary Lie bracket
of vector fields. Obviously it is linear, skew-symmetric and fulfills
the Jacobi identity.

\medskip

We will now intoduce noncommutative forms.
If we have a map \( \delta  \)
we have seen that there is a natural Lie-algebra structure (\ref{lie_delta})
over the space of derivations of the \( \star  \)-product. On this
we can easily construct the Chevalley cohomology. Further, again with
the map \( \delta  \), we can lift derivations of the Poisson structure
to derivations of the \( \star  \)-product. Therefore it should be
possible to pull back the Chevalley cohomology from the space of derivations
to vector fields. This will be done in the following.

A deformed \( k \)-form is defined to map \( k \) Poisson vector
fields to a function and has to be skew-symmetric and linear over
\( \mathbb {C} \). This is a generalization of the undeformed case,
where a form has to be linear over the algebra of functions. Functions
are defined to be \( 0 \)-forms. The space of forms \( \Omega _{\star }M \)
is now a \( \star  \)-bimodule via \begin{equation}
\label{bimodule_structure_forms}
(f\star \omega \star g)(X_{1},\dots ,X_{k})=f\star \omega (X_{1},\dots ,X_{k})\star g.
\end{equation}
As expected, the exterior differential is defined with the help of
the map \( \delta  \).\begin{equation}
\delta \omega (X_{0},\dots ,X_{k})=\sum ^{k}_{i=0}(-1)^{i}\, \delta _{X_{i}}\omega (X_{0},\dots ,\hat{X}_{i},\dots ,X_{k})\end{equation}
\begin{equation}
\label{nc_differential}
+\sum _{0\leq i<j\leq k}(-1)^{i+j}\omega ([X_{i},X_{j}]_{\star },X_{0},\dots ,\hat{X}_{i},\dots ,\hat{X}_{j},\dots ,X_{k}).\nonumber
\end{equation}
With the properties of \( \delta  \) and \( [,]_{\star } \) it follows
that \begin{equation}
\delta ^{2}\omega =0.\end{equation}
To be more explicit we give formulas for a function \( f \), a one
form \( A \) and a two form \( F \)\begin{eqnarray}
\delta f(X) & = & \delta _{X}f, \\
\delta A(X,Y) & = & \delta _{X}A_{Y}-\delta _{Y}A_{X}-A_{[X,Y]_{\star }}, \\
\delta F(X,Y,Z) & = & \delta _{X}F_{Y,Z}-\delta _{Y}F_{X,Z}+\delta _{Z}F_{X,Y},\\
 &  & -F_{[X,Y]_{\star },Z}+F_{[X,Z]_{\star },Y}-F_{[Y,Z]_{\star },X}.\nonumber 
\end{eqnarray}
A wedge product may be defined \begin{equation}
\omega _{1}\wedge \omega _{2}(X_{1},\dots ,X_{p+q})=\frac{1}{p!q!}\sum _{I,J}\varepsilon (I,J)\, \omega _{1}(X_{i_{1}},\dots ,X_{i_{p}})\star \omega _{2}(X_{j_{1}},\dots ,X_{j_{q}})\end{equation}
where \( (I,J) \) is a partition of \( (1,\dots ,p+q) \) and \( \varepsilon (I,J) \)
is the sign of the corresponding permutation. The wedge product is
linear and associative and generalizes the bimodule structure (\ref{bimodule_structure_forms}).
We note that it is no more graded commutative. We again give some
formulas.\begin{eqnarray}
(f\wedge a)_{X} & = & f\star a_{X}, \\
(a\wedge f)_{X} & = & a_{X}\star f, \\
(a\wedge b)_{X,Y} & = & a_{X}\star b_{Y}-a_{Y}\star b_{X}. 
\end{eqnarray}
The differential (\ref{nc_differential}) fulfills the graded Leibniz
rule\begin{equation}
\delta (\omega _{1}\wedge \omega _{2})=\delta \omega _{1}\wedge \omega _{2}+(-1)^{k_{2}}\, \omega _{1}\wedge \delta \omega _{2}.\end{equation}

\medskip

Now we are able to translate noncommutative gauge theory into the
language of these forms. \( A_{X} \) is the connection one form evaluated
on the vector field \( X \). It transforms like\begin{equation}
\delta _{\alpha }A=\delta \Lambda _{\alpha }+i\Lambda _{\alpha }\wedge A-A\wedge \Lambda _{\alpha }.\end{equation}
The covariant derivative of a field is now\begin{equation}
D\Psi =\delta \Psi -iA\wedge \Psi ,\end{equation}
 and the field strength becomes\begin{equation}
F=DF=\delta A-iA\wedge A.\end{equation}
One easily can show that the field strength is a covariant constant\begin{equation}
DF=\delta F-iA\wedge F=0.\end{equation}

\subsection{Seiberg-Witten gauge theory\label{swmap}}

Up to now, we could only do noncommutative gauge theory for gauge groups \(U(n)\) (see (\ref{nc_consitency_condition})). We will now show how to implement general gauge groups by using Seiberg-Witten maps \cite{Seiberg:1999vs,Jurco:2000ja}.

For general gauge groups, the commutator of two noncommutative gauge transformations no longer closes into the Lie algebra. The noncommutative gauge parameter and the noncommutative gauge potential
will therefore have to be enveloping algebra valued, but they will only depend on their
commutative counterparts, therefore preserving the right number of
degrees of freedom. These Seiberg-Witten maps \( \Lambda  \), \( \Psi  \)
and \( D \) are functionals of their classical counterparts and additionally
of the gauge potential \( a_{i} \). Their transformation properties
(\ref{nc_gt_D}) and (\ref{nc_gt_field}) should be induced by the
classical ones (\ref{com_g_field}) and (\ref{com_gt_a}) like\begin{eqnarray}
\widehat{\Lambda }_{\beta }[a]+\widehat{\delta }_{\alpha }\widehat{\Lambda }_{\beta }[a] & = & \widehat{\Lambda }_{\beta }[a+\delta _{\alpha }a],\\
\widehat{\Psi }_{\psi }[a]+\widehat{\delta }_{\alpha }\widehat{\Psi }_{\psi }[a] & = & \widehat{\Psi }_{\psi +\delta _{\alpha }\psi }[a+\delta _{\alpha }a],\\
\widehat{A}[a]+\widehat{\delta }_{\alpha }\widehat{A}[a] & = & \widehat{A}[a+\delta _{\alpha }a].
\end{eqnarray}

The Seiberg-Witten maps can be found order by order using a \( \star  \)-product
to represent the algebra on a space of functions. Translated into
this language we get for the fields \cite{Jurco:2001rq} \begin{equation}
\label{sp_gt_field}
\delta _{\alpha }\Psi _{\psi }[a]=i\Lambda _{\alpha }[a]\star \Psi _{\psi }[a].
\end{equation}

From (\ref{sp_gt_field}) we can derive a consistency condition for
the noncommuative gauge parameter \cite{Jurco:2000ja}. Insertion
into (\ref{nc_consitency_condition}) and the use of (\ref{com_consitency_condition})
yields\begin{equation}
\label{st_consitency_condition}
i\delta _{\alpha }\Lambda _{\beta }-i\delta _{\beta }\Lambda _{\alpha }+[\Lambda _{\alpha }\stackrel{\star }{,}\Lambda _{\beta }]=i\Lambda _{-i[\alpha ,\beta ]}.
\end{equation}
The transformation law for the covariantizer is now\begin{equation}
\label{sp_gt_cov}
\delta _{\alpha }(D[a](f))=i[\Lambda _{\alpha }[a]\stackrel{\star }{,}D[a](f)].
\end{equation}
The Seiberg-Witten-map can be easily extended to the derivations \(\delta_X\) of the
\( \star  \)-product. The noncommutative covariant derivative \(D_X [a]\) can be written with the help of a noncommutative gauge potential \(A_X [a]\) now depending both on the commutative gauge potential \(a\) and the vectorfield \(X\) 
\begin{equation}
\label{sp_gt_cov_derivative}
D_{X}[a]\Psi _{\psi }[a]=\delta _{X}\Psi _{\psi }[a]-iA_{X}[a]\star \Psi _{\psi }[a].
\end{equation}
It follows that the gauge potential has to transform like\begin{equation}
\label{sp_gt_gauge_pot}
\delta _{\alpha }A_{X}[a]=\delta _{X}\Lambda _{\alpha }[a]+i[\Lambda _{\alpha }[a]\stackrel{\star }{,}A_{X}[a]].
\end{equation}
We will give explicit formulas for the Seiberg-Witten maps in the chapters \ref{swmap_weyl} and \ref{swmap_form}.

\subsection{\label{Generalization to NC geometry}Gauge theory on curved noncommutative spacetime}

We are now ready to formulate gauge theory on a curved
noncommutative space, i.e. a noncommutative space with a Poisson structure
that is compatible with a frame \( e_{a} \). Later on we will propose
a method how to find such frames commuting with the Poisson structure
in the context of quantum spaces.

With the derivation \( \delta _{X} \), the covariant derivative of a field and the gauge potential now read \begin{equation}
\label{x_derivative}
D_{X}\Psi _{\psi }=\delta _{X}\Psi _{\psi }-iA_{X}\star \Psi _{\psi }.
\end{equation}
With this, a field strength may be defined as \begin{equation}
\label{xy_field_strength}
iF_{X,Y}=[D_{X}\stackrel{\star }{,}D_{Y}]-D_{[X,Y]_{\star }}.
\end{equation}
The properties of \( \delta _{\cdot } \) and \( [ \, \cdot \, , \, \cdot \,]_{\star } \)
ensure that this is really a function and not a polydifferential operator.

For a curved noncommutative space,
we can now evaluate the noncommutative covariant derivative (\ref{x_derivative})
and field strength (\ref{xy_field_strength}) on the frame \( e_{a} \) \begin{equation}
D_{a}\Phi =D_{e_a}\Phi =\delta _{e_{a}}\Phi -iA_{e_{a}}\star \Phi ,\end{equation}
\begin{equation}
F_{ab}=F(e_{a},e_{b}).\end{equation}

To write down a gauge invariant action we further need a trace, i.
e. a functional from the algebra to the complex numbers. Again the
\( \star  \)-product will be a useful tool. For a large class of
\( \star  \)-products there exist measure functions \( \Omega  \)
so that \begin{equation}
\int f\star g=\int d^{n}x\, \Omega (f\star g)\end{equation}
and \begin{equation}
\int f\star g=\int g\star f.\end{equation}
 Obviously up to first order \( \Omega  \) has to fulfill \cite{Calmet:2003jv} \begin{equation}
\partial _{\mu }(\Omega \pi ^{\mu \nu })=0.\end{equation}
 It is known \cite{Felder:2000hy} that there is always a \( \star  \)-product
for which this equation holds up to all orders.

Using the measure function and our noncommutative versions of field
strength and covariant derivative we end up with the following action
\begin{equation}\label{gen_action}
\mathcal{S}=\int d^{n}x\, \Omega \, (-\frac{1}{4}\eta ^{ab}\eta ^{cd}F_{ac}\star F_{bd}+\eta ^{ab}D_{a}\overline{\Phi }\star D_{b}\Phi -m^{2}\overline{\Phi }\star \Phi ).\end{equation}
 By construction this action is invariant under noncommutative gauge
transformations\begin{equation}
\delta _{\alpha }S=0.\end{equation}
Its classical limit is\begin{equation}
\mathcal{S}\rightarrow \int d^{n}x\, \Omega \, (-\frac{1}{4}g^{\alpha \beta }g^{\gamma \delta }f_{\alpha \gamma }f_{\beta \delta }+g^{\alpha \beta }D_{\alpha }\bar{\phi }D_{\beta }\phi -m^{2}\bar{\phi }\phi ),\end{equation}
with \( g_{\alpha \beta } \) the metric induced by the frame. In
our example (see \ref{example1} and \ref{example2}), we will further
have \( \Omega =\sqrt{g} \) and the interpretation of the classical
limit is obvious.

\medskip

We will now propose a method how to find Poisson structures and compatible
frames. On several quantum spaces deformed derivations have been constructed \cite{Wess:1991vh,Lorek:1997eh,Cerchiai:1998ef}.
In most cases the deformed Leibniz rule may be written in the following
form\begin{equation}
\hat{\partial }_{\mu }(\hat{f}\hat{g})=\hat{\partial }_{\mu }\hat{f}\hat{g}+\hat{T}_{\mu }{}^{\nu }(\hat{f})\hat{\partial }_{\nu }\hat{g},\end{equation}
where \( \hat{T} \) is an algebra morphism from the quantum space
to its matrix ring\begin{equation}
\hat{T}_{\mu }{}^{\nu }(\hat{f}\hat{g})=\hat{T}_{\mu }{}^{\alpha }(\hat{f})\hat{T}_{\alpha }{}^{\nu }(\hat{g}).\end{equation}
Again in some cases it is possible to implement this morphism with
some kind of inner morphism\begin{equation}
\hat{T}_{\mu }{}^{\nu }(\hat{f})=\hat{e}_{\mu }{}^{a}\hat{f}\hat{e}_{a}{}^{\nu },\end{equation}
where \( \hat{e}_{a}{}^{\mu } \) is an invertible matrix with entries
from the quantum space. If we define\begin{equation}
\hat{e}_{a}=\hat{e}_{a}{}^{\mu }\hat{\partial }_{\mu },\end{equation}
the \( \hat{e}_{a} \) are derivations\begin{equation}
\hat{e}_{a}(\hat{f}\hat{g})=\hat{e}_{a}(\hat{f})\hat{g}+\hat{f}\hat{e}_{a}(\hat{g}).\end{equation}
The dual formulation of this with covariant differential calculi on
quantum spaces is the formalism with commuting frames investigated
for example in \cite{Dimakis:1996,Madore:1999bi,Cerchiai:2000qi,Madore:2000aq}.
There one can additionally find how our formalism fits into the language
of Connes' spectral triples. 

We can now represent the quantum space with the help of a \( \star  \)-product.
For example we can use the Weyl ordered \( \star  \)-product we will
construct in section \ref{wely_star_construction}. Further we can
calculate the action of the operators \( \hat{e}_{a} \) on functions.
Since these are now derivations of a \( \star  \)-product, there
necessarily exist Poisson vector fields \( e_{a} \) with \begin{equation}
\delta _{e_{a}}=\hat{e}_{a}.\end{equation}

\subsection{\label{example1}Example: \protect\( SO_{a}(n)\protect \)}

In this section will examine a quantum space introduced in \cite{Majid:1994cy}.
Since we are using a \( n \)-dimensional generalisation we will simply
call it \( SO_{a}(n) \) covariant quantum space. The relations of
this quantum space are\begin{equation}
[\widehat{x}^{0},\widehat{x}^{i}]=ia\widehat{x}^{i}\; \; \; \mbox {for}\; \; \; i\neq 0,\end{equation}
 with \( a \) a real number. In the following of the example Greek
indices will run from \( 0 \) to \( n-1 \), whereas Latin indices
will run from \( 1 \) to \( n-1 \). The deformed derivations commute
and act like\begin{eqnarray}
\hat{\partial }_{o}\hat{x}^{0} & = & 1+\hat{x}^{0}\hat{\partial }_{o},\nonumber \\
\hat{\partial }_{o}\hat{x}^{i} & = & \hat{x}^{i}\hat{\partial }_{o},\\
\hat{\partial }_{i}\hat{x}^{j} & = & \delta _{i}^{j}+\hat{x}^{j}\hat{\partial }_{i},\nonumber \\
\hat{\partial }_{i}\hat{x}^{0} & = & (\hat{x}^{0}+ia)\hat{\partial }_{i},\nonumber 
\end{eqnarray}
If we define \( \hat{\rho }=\sqrt{\sum _{i}(\hat{x}^{i})^{2}} \)
and assume that it is invertible then\begin{eqnarray}
\hat{e}_{o} & = & \hat{\partial }_{0},\\
\hat{e}_{i} & = & \hat{\rho }\hat{\partial }_{i}\nonumber
\end{eqnarray}
is a frame on the quantum space. The classical limit of this is obviously\begin{eqnarray}
e_{o} & = & \partial _{0},\\
e_{i} & = & \rho \partial _{i}\nonumber 
\end{eqnarray}
and the classical metric becomes\begin{equation}
g=(dx^{0})^{2}+\rho ^{-2}((dx^{1})^{2}+\cdots +(dx^{n-1})^{2}).\end{equation}
We know that we can write\begin{equation}
(dx^{1})^{2}+\cdots +(dx^{n-1})^{2}=d\rho ^{2}+\rho ^{2}d\Omega ^{2}_{n-2},\end{equation}
where \( d\Omega ^{2}_{n-2} \) is the metric of the \( n-2 \) dimensional
sphere. Therefore in this new coordinate system \begin{equation}
g=(dx^{0})^{2}+(d\ln \rho )^{2}+d\Omega ^{2}_{n-2}\end{equation}
and we see that the classical space is a cross product of two dimensional
Euclidean space and a \( n-2 \)-sphere. Therefore it is a space of
constant non vanishing curvature. Further the measure function is\begin{equation}
\sqrt{\det g}=\rho ^{-(n-1)}.\end{equation}

We will continue this example at the end of section 3, where we will
have explicit formulas for the \( \star  \)-product.

\section{Weyl-ordered \protect\( \star \protect \)-product}

To pursue our investigation further, we will have to use a specific \(\star\)-product and construct explicit expressions for the terms that enter into the action (\ref{gen_action}).  

For the case of constant Poisson structure, one usually uses the Moyal-Weyl \(\star\)-product, corresponding to symmetric ordering of the generators of the noncommutative algebra. This procedure of generating a \(\star\)-product by an ordering prescription can be applied to more general algebras, too.

In this section we will therfore use a \( \star  \)-product generated
by symmetric ordering of the generators of a noncommutative algebra,
the Weyl-ordered \( \star  \)-product. The algebra of functions equipped
with the Weyl-ordered \( \star  \)-product is isomorphic by construction
to the noncommutative algebra it is based on. 

We will present a general formula for the Weyl-ordered \( \star  \)-product
up to second order. We will then calculate the derivations to this
\( \star  \)-product and the Seiberg-Witten maps for all relevant quantities.

\subsection{\label{wely_star_construction}Construction}

We start with an algebra generated by \( N \) elements \( \hat{x}^{i} \)
and relations

\begin{equation}
[\hat{x}^{i},\hat{x}^{j}]=\hat{c}^{ij}(\hat{x}).\end{equation}
For such an algebra we will calculate a \( \star  \)-product up to
second order. Let \begin{equation}
f(p)=\int d^{n}x\, f(x)e^{ip_{i}x^{i}}\end{equation}
 be the Fourier transform of \( f \). Then the Weyl ordered operator
associated to \( f \) is defined by\begin{equation}
W(f)=\int \frac{d^{n}p}{(2\pi )^{n}}f(p)e^{-ip_{i}\hat{x}^{i}}\end{equation}
 (see e. g. \cite{Madore:2000en}) . Every monomial of coordinate
functions is mapped to the corresponding Weyl ordered monomial of
the algebra. We note that \begin{equation}
W(e^{iq_{i}x^{i}})=e^{iq_{i}\hat{x}^{i}}.\end{equation}
The Weyl ordered \( \star  \)-product is defined by the equation\begin{equation}
W(f\star g)=W(f)W(g).\end{equation}
 If we insert the Fourier transforms of \( f \) and \( g \) we get\begin{equation}
f\star g=\int \frac{d^{n}k}{(2\pi )^{n}}\int \frac{d^{n}p}{(2\pi )^{n}}f(k)g(p)\, W^{-1}(e^{-ik_{i}\hat{x}^{i}}e^{-ip_{i}\hat{x}^{i}}).\end{equation}
We are therefore able to write down the \( \star  \)-product of the
two functions if we know the form of the last expression. For this
we expand it in terms of commutators. We use\begin{equation}
e^{\hat{A}}e^{\hat{B}}=e^{\hat{A}+\hat{B}}R(\hat{A},\hat{B})\end{equation}
with

\begin{eqnarray}
R(\hat{A},\hat{B}) & = & 1+\frac{1}{2}[\hat{A},\hat{B}] \\
 & - & \frac{1}{6}[\hat{A}+2\hat{B},[\hat{A},\hat{B}]]+\frac{1}{8}[\hat{A},\hat{B}][\hat{A},\hat{B}]+\mathcal{O}(3).\nonumber
\end{eqnarray}
If we set \( \hat{A}=-ik_{i}\hat{x}^{i} \) and \( \hat{B}=-ip_{i}\hat{x}^{i} \)
the above-mentioned expression becomes\begin{equation}
W^{-1}(e^{-ik_{i}\hat{x}^{i}}e^{-ip_{i}\hat{x}^{i}})=\nonumber \end{equation}
 \begin{eqnarray}
 &  & e^{-i(k_{i}+p_{i})x^{i}}+\frac{1}{2}(-ik_{i})(-ip_{j})W^{-1}(e^{-i(k_{i}+p_{i})\hat{x}^{i}}[\hat{x}^{i},\hat{x}^{j}])\nonumber \\
 &  & -\frac{1}{6}(-i)(k_{m}+2p_{m})(-ik_{i})(-ip_{j})W^{-1}(e^{-i(k_{i}+p_{i})\hat{x}^{i}}[[\hat{x}^{m},[\hat{x}^{i},\hat{x}^{j}]])\\
 &  & +\frac{1}{8}(-ik_{m})(-ip_{n})(-ik_{i})(-ip_{j})W^{-1}(e^{-i(k_{i}+p_{i})\hat{x}^{i}}[\hat{x}^{m},\hat{x}^{n}][\hat{x}^{i},\hat{x}^{j}])\nonumber \\
 &  & +\mathcal{O}(3).\nonumber 
\end{eqnarray}
If we assume that the commutators of the generators are written in
Weyl ordered form\begin{equation}
\hat{c}^{ij}=W(c^{ij}),\end{equation}
we see that\begin{equation}
[\hat{x}^{m},[\hat{x}^{i},\hat{x}^{j}]]=W(c^{ml}\partial _{l}c^{ij})+\mathcal{O}(3),\end{equation}
\begin{equation}
[\hat{x}^{m},\hat{x}^{n}][\hat{x}^{i},\hat{x}^{j}]=W(c^{mn}c^{ij})+\mathcal{O}(3).\end{equation}
Further we can derive\begin{eqnarray}
W^{-1}(e^{iq_{i}\hat{x}^{i}}W(f)) & = & W^{-1}\left( \int \frac{d^{n}p}{(2\pi )^{n}}f(p)e^{-i(q_{i}+p_{i})\hat{x}^{i}}R(-iq_{i}\hat{x}^{i},-ip_{i}\hat{x}^{i})\right) \nonumber \\
 & = & e^{-iq_{i}x^{i}}\left( f+\frac{1}{2}(-iq_{i})c^{ij}\partial _{j}f\right) +\mathcal{O}(2).
\end{eqnarray}
Putting all this together yields

\begin{eqnarray}
W^{-1}(e^{-ik_{i}\hat{x}^{i}}e^{-ip_{i}\hat{x}^{i}}) & = & e^{-i(k_{i}+p_{i})x^{i}}\left( 1+\frac{1}{2}c^{ij}(-ik_{i})(-ip_{j})\right.  \\
 & + & \frac{1}{8}c^{mn}c^{ij}(-ik_{m})(-ip_{n})(-ik_{i})(-ip_{j})\nonumber \\
 & + & \left. \frac{1}{12}c^{ml}\partial _{l}c^{ij}(-i)(k_{m}-p_{m})(-ik_{i})(-ip_{j})\right) \nonumber \\
 & + & \mathcal{O}(3),\nonumber 
\end{eqnarray}
and we can write down the Weyl ordered \( \star  \)-product up to
second order for an arbitrary algebra\begin{eqnarray}
f\star g & = & fg+\frac{1}{2}c^{ij}\partial _{i}f\partial _{j}g\label{wo_product} \\
 & + & \frac{1}{8}c^{mn}c^{ij}\partial _{m}\partial _{i}f\partial _{n}\partial _{j}g\nonumber \\
 & + & \frac{1}{12}c^{ml}\partial _{l}c^{ij}(\partial _{m}\partial _{i}f\partial _{j}g-\partial _{i}f\partial _{m}\partial _{j}g)+\mathcal{O}(3).\nonumber 
\end{eqnarray}

Let us collect some properties of the just calculated \( \star  \)-product.
First\begin{equation}
[x^{i}\stackrel{\star }{,}x^{j}]=c^{ij}\end{equation}
is the Weyl ordered commutator of the algebra. Further, if there is
a conjugation on the algebra and if we assume that the noncommutative
coordinates are real \( \overline{\hat{x}^{i}}=\hat{x}^{i} \), then
the Weyl ordered monomials are real, too. This is also true for the
monomials of the commutative coordinate functions. Therefore this
\( \star  \)-product respects the ordinary complex conjugation\begin{equation}
\overline{f\star g}=\overline{g}\star \overline{f}.\end{equation}
On the level of the Poisson tensor this means\begin{equation}
\overline{c^{ij}}=-c^{ij}.\end{equation}
If we have a measure function \( \Omega  \) with \( \partial _{i}(\Omega c^{ij})=0 \),
then \begin{equation}
\int d^{n}x\, \Omega \, f\star g=\int d^{n}x\, \Omega \, g\star f+\mathcal{O}(3).\end{equation}

\subsection{\label{weyl derivations}Derivations}
We now want to calculate the derivations \( \delta_X  \) of the Weyl-ordered \(\star\)-product (\ref{wo_product}) from the derivations \(X\) of the Poisson structure \(c^{ij}\) up to second order. We assume
that \( \delta _{X} \) can be expanded in the following way\begin{equation}
\delta _{X}=X^{i}\partial _{i}+\delta ^{ij}_{X}\partial _{i}\partial _{j}+\delta _{X}^{ijk}\partial _{i}\partial _{j}\partial _{k}+\cdots .\end{equation}
 Expanding the equation \begin{equation}
\delta _{X}(f\star g)=\delta _{X}(f)\star g+f\star \delta _{X}(g)\end{equation}
 order by order and using \( [X,c]_{S}=0 \) we find that\begin{eqnarray}
\delta _{X} & = & X^{i}\partial _{i}-\frac{1}{12}c^{lk}\partial _{k}c^{im}\partial _{l}\partial _{m}X^{j}\partial _{i}\partial _{j} \\
 &  & +\frac{1}{24}c^{lk}c^{im}\partial _{l}\partial _{i}X^{j}\partial _{k}\partial _{m}\partial _{j}+\mathcal{O}(3).\nonumber
\end{eqnarray}
For \( [ \, \cdot \, , \, \cdot \,]_{\star }\) we simply calculate \( [\delta _{X},\delta _{Y}] \)
and get\begin{eqnarray}
[X,Y]_{\star } & = & [X,Y]_{L} \\
 &  & -\frac{1}{12}(c^{lk}\partial _{k}c^{im}\partial _{l}\partial _{m}X^{j}\partial _{i}\partial _{j}Y^{n}-c^{lk}\partial _{k}c^{im}\partial _{l}\partial _{m}Y^{j}\partial _{i}\partial _{j}X^{n})\partial _{n}\nonumber \\
 &  & +\frac{1}{24}(c^{lk}c^{im}\partial _{l}\partial _{i}X^{j}\partial _{k}\partial _{m}\partial _{j}Y^{n}-c^{lk}c^{im}\partial _{l}\partial _{i}Y^{j}\partial _{k}\partial _{m}\partial _{j}X^{n})\partial _{n} \nonumber \\
 &  & +\mathcal{O}(3).\nonumber 
\end{eqnarray}

\subsection{Explicit formulas for the Seiberg-Witten map \label{swmap_weyl}}

We will now present a consistent solution for the Seiberg-Witten-maps
up to second order for the Weyl ordered \( \star  \)-product and
non-abelian classical gauge transformations. The solutions have been
chosen in such a way that they reproduce the ones obtained in \cite{Jurco:2001rq}
for the constant case.

The solution for the gauge transformations is obtained by solving
the consistency condition (\ref{st_consitency_condition}) order by
order\begin{eqnarray}
\Lambda _{\alpha }[a] & = & \alpha -\frac{i}{4}c^{ij}\{\partial _{i}\alpha ,a_{j}\}\nonumber \\
 &  & +\frac{1}{32}c^{ij}c^{kl}\Big {(}4\{\partial _{i}\alpha ,\{a_{k},\partial _{l}a_{j}\}\}-2i[\partial _{i}\partial _{k}\alpha ,\partial _{j}a_{l}]\nonumber \\
 &  & \; \; \; \; \; \; +2[\partial _{j}a_{l},[\partial _{i}\alpha ,a_{k}]]-2i[[a_{j},a_{l}],[\partial _{i}\alpha ,a_{k}]]\\
 &  & \; \; \; \; \; \; +i\{\partial _{i}\alpha ,\{a_{k},[a_{j},a_{l}]\}\}+\{a_{j},\{a_{l},[\partial _{i}\alpha ,a_{k}]\}\}\Big {)}\nonumber \\
 &  & +\frac{1}{24}c^{kl}\partial _{l}c^{ij}\Big {(}\{\partial _{i}\alpha ,\{a_{k},a_{j}\}\}-2i[\partial _{i}\partial _{k}\alpha ,a_{j}]\Big {)}+\mathcal{O}(3).\nonumber 
\end{eqnarray}
In the same way a solution for the field is obtained by solving
equation (\ref{sp_gt_field})\begin{eqnarray}
\Psi _{\psi }[a] & = & \psi +\frac{1}{4}c^{ij}\Big {(}2ia_{i}\partial _{j}\psi +a_{i}a_{j}\psi \Big {)}\nonumber \\
 &  & +\frac{1}{32}c^{ij}c^{kl}\Big {(}4i\partial _{i}a_{k}\partial _{j}\partial _{l}\psi -4a_{i}a_{k}\partial _{j}\partial _{l}\psi -8a_{i}\partial _{j}a_{k}\partial _{l}\psi \nonumber \\
 &  & \; \; \; \; \; \; +4a_{i}\partial _{k}a_{j}\partial _{l}\psi +4ia_{i}a_{j}a_{k}\partial _{l}\psi -4ia_{k}a_{j}a_{i}\partial _{l}\psi \nonumber \\
 &  & \; \; \; \; \; \; +4ia_{j}a_{k}a_{i}\partial _{l}\psi -4\partial _{j}a_{k}a_{i}\partial _{l}\psi +2\partial _{i}a_{k}\partial _{j}a_{l}\psi \nonumber \\
 &  & \; \; \; \; \; \; -4ia_{i}a_{l}\partial _{k}a_{j}\psi -4ia_{i}\partial _{k}a_{j}a_{l}\psi +4ia_{i}\partial _{j}a_{k}a_{l}\psi \\
 &  & \; \; \; \; \; \; -3a_{i}a_{j}a_{l}a_{k}\psi -4a_{i}a_{k}a_{j}a_{l}\psi -2a_{i}a_{l}a_{k}a_{j}\psi \Big {)}\nonumber \\
 &  & +\frac{1}{24}c^{kl}\partial _{l}c^{ij}\Big {(}2ia_{j}\partial _{k}\partial _{i}\psi +2i\partial _{k}a_{i}\partial _{j}\psi +2\partial _{k}a_{i}a_{j}\psi \nonumber \\
 &  & \; \; \; \; \; \; -a_{k}a_{i}\partial _{j}\psi -3a_{i}a_{k}\partial _{j}\psi -2ia_{j}a_{k}a_{i}\psi \Big {)}+\mathcal{O}(3).\nonumber 
\end{eqnarray}
From (\ref{sp_gt_cov}) the covariantizer becomes

\begin{equation}
\begin{array}{rcl}
D[a](f) & = & f+ic^{ij}a_{i}\partial _{j}f \\
 &  & +\frac{1}{4}c^{ij}c^{kl}\Big {(}-2\{a_{i},\partial _{j}a_{k}\}\partial _{l}f+\{a_{i},\partial _{k}a_{j}\}\partial _{l}f \\
 &  & \; \; \; \; \; \; +i\{a_{i},[a_{j},a_{k}]\}\partial _{l}f-\{a_{i},a_{k}\}\partial _{j}\partial _{l}f\Big {)} \\
 &  & +\frac{1}{4}c^{il}\partial _{l}c^{jk}\{a_{i},a_{k}\}\partial _{j}f+\mathcal{O}(3).
\end{array}\end{equation}
With (\ref{sp_gt_gauge_pot}) the NC gauge potential is 
\begin{eqnarray}
A_{X} & = & X^{n}a_{n}+\frac{i}{4}c^{kl}X^{n}\{a_{k},\partial _{l}a_{n}+f_{ln}\}+\frac{i}{4}c^{kl}\partial _{l}X^{n}\{a_{k},a_{n}\}\nonumber \\
 &  & +\frac{1}{32}c^{kl}c^{ij}X^{n}\Big {(}-4i[\partial _{k}\partial _{i}a_{n},\partial _{l}a_{j}]+2i[\partial _{k}\partial _{n}a_{i},\partial _{l}a_{j}]-4\{a_{k},\{a_{i},\partial _{j}f_{ln}\}\}\nonumber \\
 &  & \; \; \; \; \; \; \; \; \; -2[[\partial _{k}a_{i},a_{n}],\partial _{l}a_{j}]+4\{\partial _{l}a_{n},\{\partial _{i}a_{k},a_{j}\}\}-4\{a_{k},\{f_{li},f_{jn}\}\}\nonumber \\
 &  & \; \; \; \; \; \; \; \; \; +i\{\partial _{n}a_{j},\{a_{l},[a_{i},a_{k}]\}\}+i\{a_{i},\{a_{k},[\partial _{n}a_{j},a_{l}]\}\}-4i[[a_{i},a_{l}],[a_{k},\partial _{j}a_{n}]]\nonumber \\
 &  & \; \; \; \; \; \; \; \; \; +2i[[a_{i},a_{l}],[a_{k},\partial _{n}a_{j}]]+\{a_{i},\{a_{k},[a_{l},[a_{j},a_{n}]]\}\}\nonumber \\
 &  & \; \; \; \; \; \; \; \; \; -\{a_{k},\{[a_{l},a_{i}],[a_{j},a_{n}]\}\}-[[a_{i},a_{l}],[a_{k},[a_{j},a_{n}]]]\Big {)}\nonumber \\
 &  & +\frac{1}{32}c^{kl}c^{ij}\partial _{j}X^{n}\Big {(}2i[\partial _{k}a_{i},\partial _{l}a_{n}]+2i[\partial _{i}a_{k},\partial _{l}a_{n}]+2i[\partial _{i}a_{k},\partial _{l}a_{n}-\partial _{n}a_{l}]\nonumber \\
 &  & \; \; \; \; \; \; \; \; \; +4\{a_{n},\{a_{l},\partial _{k}a_{i}\}\}+4\{a_{k},\{a_{i},\partial _{n}a_{l}-\partial _{l}a_{n}\}\}-2i\{a_{k},\{a_{i},[a_{n},a_{l}]\}\}\nonumber \\
 &  & \; \; \; \; \; \; \; \; \; +i\{a_{i},\{a_{l},[a_{n},a_{k}]\}\}+i\{a_{n},\{a_{l},[a_{i},a_{k}]\}\}\Big {)} \\
 &  & +\frac{1}{24}c^{kl}c^{ij}\partial _{l}\partial _{j}X^{n}\Big {(}\partial _{i}\partial _{k}a_{n}-2i[a_{i},\partial _{k}a_{n}]-\{a_{n},\{a_{k},a_{i}\}\}\Big {)}\nonumber \\
 &  & +\frac{1}{24}c^{kl}\partial _{l}c^{ij}X^{n}\Big {(}2i[a_{j},\partial _{k}\partial _{i}a_{n}]+2i[\partial _{k}a_{i},f_{jn}]\nonumber \\
 &  & \; \; \; \; \; \; \; \; \; -\{\partial _{j}a_{n},\{a_{k},a_{i}\}\}+2\{a_{i},\{a_{k},f_{nj}\}\}\Big {)}\nonumber \\
 &  & +\frac{1}{24}c^{kl}\partial _{l}c^{ij}\partial _{j}X^{n}\Big {(}-4i[a_{i},\partial _{k}a_{n}]+2i[a_{k},\partial _{i}a_{n}]-\{a_{n},\{a_{k},a_{i}\}\}\Big {)}\nonumber \\
 &  & -\frac{1}{12}c^{kl}\partial _{l}c^{ij}\partial _{j}\partial _{k}X^{n}\partial _{i}a_{n}+\mathcal{O}(3).\nonumber 
\end{eqnarray}
The resulting field strength is\begin{eqnarray}
F_{ab}=F(X_{a},X_{b}) & = & X^{k}_{a}X^{l}_{b}f_{kl}+\frac{i}{2}c^{ij}\{a_{i},\partial _{j}(X^{k}_{a}X^{l}_{b}f_{kl})\}\\
 &  & +\frac{i}{2}c^{ij}X^{k}_{a}X^{l}_{b}\{f_{jl},f_{ki}\}+\frac{1}{4}c^{ij}X^{k}_{a}X^{l}_{b}\{a_{i},[a_{j},f_{kl}]\}+\mathcal{O}(2).\nonumber 
\end{eqnarray}
The covariant derivative is\begin{eqnarray}
D_{a}\Phi =D_{X_{a}}\Phi  & = & \delta _{X_{a}}\Phi -iA_{X_{a}}\star \Phi \\
 & = & X_{a}^{k}D_{k}\phi +\frac{i}{2}X_{a}^{k}f_{ki}c^{ij}D_{j}\phi\nonumber  \\
 &  & +\frac{i}{2}c^{ij}a_{i}\partial _{j}(X_{a}^{k}D_{k}\phi )+\frac{1}{4}c^{ij}a_{i}a_{j}X_{a}^{k}D_{k}\phi +\mathcal{O}(2).\nonumber 
\end{eqnarray}
Using partial integration and the trace property of the integral,
i.e. \( \partial _{\mu }(\Omega c^{\mu \nu })=0 \), we can calculate 

\begin{eqnarray}
\widetilde{S}_{gauge} & = & \int d^{n}x\, \Omega \, \eta ^{ab}\eta ^{cd}F_{ac}\star F_{bd} \\
 & = & \int d^{n}x\, \Omega \, \eta ^{ab}\eta ^{cd}X^{\mu }_{a}X^{\nu }_{c}X^{\rho }_{b}X^{\sigma }_{d}f_{\mu \nu }f_{\rho \sigma }\nonumber \\
 &  & +\int d^{4}x\, \Omega \, \eta ^{ab}\eta ^{cd}\Big {(}\frac{i}{4}c^{ij}[a_{i},[\partial _{j}(X^{\mu }_{a}X^{\nu }_{c}f_{\mu \nu }),X^{\rho }_{b}X^{\sigma }_{d}f_{\rho \sigma }]]\nonumber \\
 &  & +\frac{i}{8}c^{ij}X^{\mu }_{a}X^{\nu }_{c}X^{\rho }_{b}X^{\sigma }_{d}\{f_{\mu \nu },\{f_{ij},f_{\rho \sigma }\}\}-\frac{1}{4}c^{ij}X^{\mu }_{a}X^{\nu }_{c}X^{\rho }_{b}X^{\sigma }_{d}[f_{\mu \nu }a_{i},f_{\rho \sigma }a_{j}]\nonumber \\
 &  & -\frac{1}{4}c^{ij}X^{\mu }_{a}X^{\nu }_{c}X^{\rho }_{b}X^{\sigma }_{d}[a_{i}f_{\mu \nu },a_{j}f_{\rho \sigma }]+\frac{i}{2}c^{ij}X^{\mu }_{a}X^{\nu }_{c}X^{\rho }_{b}X^{\sigma }_{d}\{f_{\mu \nu },\{f_{j\sigma },f_{\rho i}\}\}\Big {)}+\mathcal{O}(2).\nonumber 
\end{eqnarray}
Therefore the action for the gauge particles is\begin{eqnarray}
S_{gauge} & = & -\frac{1}{4}tr(\widetilde{S}_{gauge})\\
 & = & \int d^{n}x\, \Omega \, \eta ^{ab}\eta ^{cd}X^{\mu }_{a}X^{\nu }_{c}X^{\rho }_{b}X^{\sigma }_{d}\Big {(}-\frac{1}{4}tr(f_{\mu \nu }f_{\rho \sigma })\nonumber \\
 &  & -\frac{i}{8}c^{ij}tr(f_{ij}f_{\mu \nu }f_{\rho \sigma })-\frac{i}{2}c^{ij}tr(f_{\mu \nu }f_{j\sigma }f_{\rho i})\Big {)}+\mathcal{O}(2).\nonumber 
\end{eqnarray}
With \( \overline{c}^{ij}=-c^{ij} \) we get

\begin{eqnarray}
S_{scalar} & = & \int d^{n}x\, \Omega \, \eta ^{ab}\overline{D_{a}\Phi }\star D_{b}\Phi \\
 & = & \int d^{n}x\, \Omega \, \eta ^{ab}\Big {(}\overline{X}_{a}^{\mu }X^{\nu }_{b}\overline{D_{\mu }\phi }D_{\nu }\phi\nonumber  \\
 &  & +\frac{i}{2}c^{ij}\overline{X}^{\mu }_{a}X^{\nu }_{b}\overline{D_{\mu }\phi }f_{\nu i}D_{j}\phi +\frac{i}{2}c^{ij}\overline{X}^{\mu }_{a}X^{\nu }_{b}\overline{D_{j}\phi }f_{\nu i}D_{\mu }\phi\nonumber  \\
 &  & +\frac{i}{2}c^{ij}\overline{X}^{\mu }_{a}X^{\nu }_{b}\overline{D_{\mu }\phi }f_{ij}D_{\nu }\phi \Big {)}+\mathcal{O}(2)\nonumber 
\end{eqnarray}
for the scalar fields.

\subsection{\label{example2}Example: \protect\( SO_{a}(n)\protect \)}

Now we continue our example from chapter \ref{example1}. It is easy to see that
the Poisson tensor corresponding to the algebra is\begin{equation}
c^{\mu \nu }=ia\delta ^{\mu }_{0}\delta _{i}^{\nu }x^{i}-ia\delta ^{\nu }_{0}\delta _{i}^{\mu }x^{i}.\end{equation}
Since we are dealing here with the case of a Lie algebra we surely
have \( W(c^{\mu \nu })=[\hat{x}^{\mu },\hat{x}^{\nu }] \). The \( \hat{x}^{i} \)
commute with each other, therefore we get with \( \rho =\sqrt{x_{i}x^{i}} \)
\begin{equation}
W(\rho )=\hat{\rho }.\end{equation}
The components of the frame (\( e_{\alpha }=X_{\alpha }{}^{\mu }\partial _{u} \))
are\begin{eqnarray}
X^{\mu }_{0} & = & \delta _{0}^{\mu },\\
X^{\mu }_{i} & = & \rho \delta ^{\mu }_{i}.\nonumber 
\end{eqnarray}
These we can plug into our solution of the Seiberg-Witten map and
get\begin{eqnarray}
\Lambda _{\lambda }[a] & = & \lambda +\frac{a}{4}x^{i}\{\partial _{0}\lambda ,a_{i}\}-\frac{a}{4}x^{i}\{\partial _{i}\lambda ,a_{0}\}+\mathcal{O}(a^{2}),\nonumber \\
\Phi _{\phi }[a] & = & \phi -\frac{a}{2}x^{i}a_{0}\partial _{i}\phi +\frac{a}{2}x^{i}a_{i}\partial _{0}\phi +\frac{ia}{4}x^{i}[a_{0},a_{i}]\phi +\mathcal{O}(a^{2}),\nonumber \\
A_{X_{0}} & = & a_{0}-\frac{a}{4}x^{i}\{a_{0},\partial _{i}a_{0}+f_{i0}\}+\frac{a}{4}x^{i}\{a_{i},\partial _{0}a_{0}\}+\mathcal{O}(a^{2}),\\
A_{X_{j}} & = & \rho a_{j}-\frac{a}{4}\rho \{a_{j},a_{0}\}-\frac{a}{4}\rho x^{i}\{a_{0},\partial _{i}a_{j}+f_{ij}\}+\frac{a}{4}\rho x^{i}\{a_{i},\partial _{0}a_{j}+f_{0j}\}+\mathcal{O}(a^{2}),\nonumber \\
\delta _{X_{\mu }} & = & X^{\nu }_{\mu }\partial _{\nu }+\mathcal{O}(a^{2}).\nonumber 
\end{eqnarray}
A measure function induced by the trace property of the integral up
to second order is\begin{equation}
\Omega =\rho ^{-(n-1)}.\end{equation}
We note that\begin{equation}
\Omega =\sqrt{g},\end{equation}
where \( g_{\alpha \beta } \) is the classical metric induced by
the noncommutative frame (see section 2). With this measure function
the actions become\begin{eqnarray}
S_{gauge} & = & -\frac{1}{2}\int d^{n}x\, \rho ^{3-n}\eta ^{00}\eta ^{ij}Tr(f_{0i}f_{0j})-\frac{1}{4}\int d^{4}x\, \rho ^{5-n}\eta ^{kl}\eta ^{ij}Tr(f_{ki}f_{lj}) \\
 &  & -\frac{a}{2}\int d^{n}x\, \rho ^{3-n}\eta ^{00}\eta ^{ij}x^{p}Tr(f_{0p}f_{0i}f_{0j})+\frac{a}{4}\int d^{4}x\, \rho ^{5-n}\eta ^{kl}\eta ^{ij}x^{p}Tr(f_{0p}f_{ki}f_{lj})\nonumber \\
 &  & -\frac{a}{2}\int d^{n}x\, \rho ^{5-n}\eta ^{kl}\eta ^{ij}x^{p}Tr(f_{jp}\{f_{ki},f_{l0}\})+\mathcal{O}(a^{2})\nonumber 
\end{eqnarray}
and\begin{eqnarray}
S_{scalar} & = & \int d^{n}x\, \rho ^{1-n}\eta ^{00}\overline{D_{0}\phi }D_{0}\phi +\int d^{n}x\, \rho ^{3-n}\eta ^{kl}\overline{D_{k}\phi }D_{l}\phi \\
 &  & -\frac{a}{2}\int d^{n}x\, \rho ^{3-n}\eta ^{kl}x^{i}\overline{D_{k}\phi }f_{l0}D_{i}\phi +\frac{a}{2}\int d^{n}x\, \rho ^{3-n}\eta ^{kl}x^{i}\overline{D_{k}\phi }f_{li}D_{0}\phi\nonumber  \\
 &  & -\frac{a}{2}\int d^{n}x\, \rho ^{3-n}\eta ^{kl}x^{i}\overline{D_{i}\phi }f_{l0}D_{k}\phi +\frac{a}{2}\int d^{n}x\, \rho ^{3-n}\eta ^{kl}x^{i}\overline{D_{0}\phi }f_{li}D_{k}\phi\nonumber  \\
 &  & -a\int d^{n}x\, \rho ^{3-n}\eta ^{kl}x^{i}\overline{D_{k}\phi }f_{0i}D_{l}\phi +\mathcal{O}(a^{2}).\nonumber 
\end{eqnarray}
In the classical limit \( a\rightarrow 0 \) the action reduces to
scalar electrodynamics on a curved background or its nonabelian generalization,
respectively.

\section{Formality}

The Weyl-ordered \(\star\)-product of the last section is very useful for explicit calculations, but these can only be done in a perturbative way order by order. Also, it is only known in general up to the second order we calculated in chapter \ref{wely_star_construction}. For closed expessions and questions of existence, Kontsevich's Formality \( \star  \)-product \cite{Kontsevich:1997vb} is the better choice. It is known to all orders and comes with a strong mathematical framework, in which
the derivations are obtained in a very natural way. 

\subsection{\label{Formality map}The Formality Map}

Kontsevich's Formality map \cite{Kontsevich:1997vb} is a very useful
tool for studying the relations between Poisson tensors and \( \star  \)-products.
To make use of the Formality map we first want to recall some definitions.
A polyvector field is a skew-symmetric tensor in the sense of differential
geometry. Every \( n \)-polyvector field \( \alpha  \) may locally
be written as\begin{equation}
\alpha =\alpha ^{i_{1}\dots i_{n}}\, \partial _{i_{1}}\wedge \dots \wedge \partial _{i_{n}}.\end{equation}
 We see that the space of polyvector fields can be endowed with a
grading \( n \). For polyvector fields there is a grading respecting
bracket that in a natural way generalizes the Lie bracket \( [ \, \cdot \, , \,
\cdot \,]_{L} \)
of two vector fields, the Schouten-Nijenhuis bracket. For an exact
definition see \ref{Schouten-Nijenhuis bracket}. If \( \pi  \) is
a Poisson tensor, the Hamiltonian vector field \( H_{f} \) for a
function \( f \) is\begin{equation}
H_{f}={[\pi ,f]_{S}}=-\pi ^{ij}\partial _{i}f\partial _{j}.\end{equation}
Note that \( [\pi ,\pi ]_{S}=0 \) is the Jakobi identity of a Poisson
tensor. \medskip

On the other hand a \( n \)-polydifferential operator is a multilinear
map that maps \( n \) functions to a function. For example, we may
write a \( 1 \)-polydifferential operator \( D \) as\begin{equation}
D(f)=D_{0}f+D^{i}_{1}\partial _{i}f+D^{ij}_{2}\partial _{i}\partial _{j}f+\dots .\end{equation}
The ordinary multiplication \( \cdot  \) is a \( 2 \)-polydifferential
operator. It maps two functions to one function. Again the number
\( n \) is a grading on the space of polydifferential operators.
Now the Gerstenhaber bracket is natural and respects the grading.
For a exact definition see \ref{Gerstenhaber bracket}. 

The Formality map is a collection of skew-symmetric multilinear maps
\( U_{n} \), \( n=0,1,\dots  \), that maps \( n \) polyvector fields
to a \( m \)-differential operator. To be more specific let \( \alpha _{1},\dots ,\alpha _{n} \)
be polyvector fields of grade \( k_{1},\dots ,k_{n} \). Then \( U_{n}(\alpha _{1},\dots ,\alpha _{n}) \)
is a polydifferential operator of grade\begin{equation}
m=2-2n+\sum _{i}k_{i}.\end{equation}
 In particular the map \( U_{1} \) is a map from a \( k \)-vectorfield
to a \( k \)-differential operator. It is defined by\begin{equation}
U_{1}(\alpha ^{i_{1}\dots i_{n}}\partial _{i_{1}}\wedge \dots \wedge \partial _{i_{n}})(f_{1},\dots ,f_{n})=\alpha ^{i_{1}\dots i_{n}}\partial _{i_{1}}f_{1}\cdot \dots \cdot \partial _{i_{n}}f_{n}.\end{equation}
The formality maps \( U_{n} \) fulfill the formality condition \cite{Kontsevich:1997vb,Arnal:2000hy}

\begin{equation}
\label{formality cond}
Q'_{1}U_{n}(\alpha _{1},\ldots ,\alpha _{n})+\frac{1}{2}\sum _{{I\sqcup J=\{1,\ldots ,n\}\atop I,J\neq \emptyset }}\epsilon (I,J)Q'_{2}(U_{|I|}(\alpha _{I}),U_{|J|}(\alpha _{J})) 
\end{equation}
\begin{equation}
=\frac{1}{2}\sum _{i\neq j}\epsilon (i,j,\ldots ,\hat{i},\ldots ,\hat{j},\ldots ,n)U_{n-1}(Q_{2}(\alpha _{i},\alpha _{j}),\alpha _{1},\ldots ,\widehat{\alpha }_{i},\ldots ,\widehat{\alpha }_{j},\ldots ,\alpha _{n}).\nonumber \end{equation}
The hats stand for omitted symbols, \( Q'_{1}(\Upsilon )=[\Upsilon ,\mu ] \)
with \( \mu  \) being ordinary multiplication and \( Q'_{2}(\Upsilon _{1},\Upsilon _{2})=(-1)^{(|\Upsilon _{1}|-1)|\Upsilon _{2}|}[\Upsilon _{1},\Upsilon _{2}]_{G} \)
with \( |\Upsilon _{s}| \) being the degree of the polydifferential
operator \( \Upsilon _{s} \), i.e. the number of functions it is
acting on. For polyvectorfields \( \alpha ^{i_{1}\ldots i_{k_{s}}}_{s}\partial _{i_{1}}\wedge \ldots \wedge \partial _{i_{k_{s}}} \)
of degree \( k_{s} \) we have \( Q_{2}(\alpha _{1},\alpha _{2})=-(-1)^{(k_{1}-1)k_{2}}[\alpha _{2},\alpha _{1}]_{S} \). 

For a bivectorfield \( \pi  \) we can now define a bidifferential
operator

\begin{equation}
\star =\sum ^{\infty }_{n=0}\frac{1}{n!}U_{n}(\pi ,\, \ldots ,\pi )\end{equation}
i.e.

\begin{equation}
f\star g=\sum ^{\infty }_{n=0}\frac{1}{n!}U_{n}(\pi ,\, \ldots ,\pi )(f,g).\end{equation}
We further define the special polydifferential operators

\begin{eqnarray}
\Phi (\alpha ) & = & \sum ^{\infty }_{n=1}\frac{1}{(n-1)!}U_{n}(\alpha ,\pi ,\, \ldots ,\pi ),\\
\Psi (\alpha _{1},\alpha _{2}) & = & \sum ^{\infty }_{n=2}\frac{1}{(n-2)!}U_{n}(\alpha _{1},\alpha _{2},\pi ,\ldots ,\pi ).
\end{eqnarray}
For \( g \) a function, \( X \) and \( Y \) vectorfields and \( \pi  \)
a bivectorfield we see that\begin{equation}
\delta _{X}=\Phi (X)\end{equation}
is a 1-differential operator and that both \( \phi (g) \) and \( \Psi (X,Y) \)
are functions. 

We now use the formality condition (\ref{formality cond}) to calculate
\begin{eqnarray}
{[\star ,\star ]_{G}} & = & \Phi ([\pi ,\pi ]_{S}),\label{comm star star} \\
{[\Phi (f),\star ]_{G}} & = & -\Phi ([f,\pi ]_{S}),\label{comm phi f star} \\
{[\delta _{X},\star ]_{G}} & = & \Phi ([X,\pi ]_{S}),\label{comm delta X star} \\
{[\delta _{X},\delta _{Y}]_{G}+[\Psi (X,Y),\star ]_{G}} & = & \delta _{[X,Y]_{S}}+\Psi ([\theta ,Y]_{S},X)-\Psi ([\theta ,X]_{S},Y),\label{comm delta X delta Y} \\
{}[\Phi (\pi ),\Phi (g)]_{G}+[\Psi (\pi ,g),\star ]_{G} & = & -\delta _{[\pi ,g]_{S}}-\Psi ([\theta ,g]_{S},\pi )-\Psi ([\theta ,\pi ]_{S},g),\\
{}[\delta _{X},\Phi (g)]_{G} & = & \phi ([X,g]_{S})-\Psi ([\theta ,g]_{S},X)-\Psi ([\theta ,X]_{S},g).\label{comm delta X Phi g} 
\end{eqnarray}
 If \( \pi  \) is Poisson, i. e. \( [\pi ,\pi ]_{S}=0 \) and if
\( X \) and \( Y \) are Poisson vector fields, i. e. \( [X,\pi ]_{S}=[Y,\pi ]_{S}=0 \),
the relations (\ref{comm star star}) to (\ref{comm delta X delta Y})
become \begin{eqnarray}
f\star (g\star h) & = & (f\star g)\star h,\nonumber \\
\delta _{H_{f}}(g) & = & -[\Phi (f)\stackrel{\star }{,}g],\nonumber \\
\delta _{X}(f\star g) & = & \delta _{X}(f)\star g+f\star \delta _{X}(g),\label{formality star derivation} \\
({[\delta _{X},\delta _{Y}]-\delta _{[X,Y]_{L}}})(g) & = & [\Psi (X,Y)\stackrel{\star }{,}g].\nonumber \label{formality star derivation} 
\end{eqnarray}
when evaluated on functions. \( [\, \cdot \, , \, \cdot \,]  \) are now ordinary brackets.
\( \star  \) defines an associative product, the Hamiltonian vector
fields are mapped to inner derivations and Poisson vector fields are
mapped to outer derivations of the \( \star  \)-product. Note that in \cite{Cornalba:2001sm} non-associative \( \star \)-products arised when curved D-branes were considered on curved backgrounds. 
 
Additionally the map \( \delta  \) preserves the bracket up to an inner derivation.
The last equation can be cast into a form we used extensively in the
definition of our deformed forms;\begin{equation}
{[\delta _{X},\delta _{Y}]}=\delta _{[X,Y]_{\star }}\end{equation}
with\begin{equation}
{[X,Y]_{\star }}=[X,Y]_{L}+H_{\Phi ^{-1}\Psi (X,Y)}.\end{equation}

\subsection{Construction of the Seiberg-Witten map \label{swmap_form}}

With the formality \( \star  \)-product and the derivations on it
we have all the key ingredients to do NC gauge theory on any Poisson
manifold. To relate the NC theory to commutative gauge theory, we
need the Seiberg-Witten maps for the formality \( \star  \)-product.
In \cite{Jurco:2000fs} and \cite{Jurco:2001my} the SW maps for the
NC gauge parameter and the Covariantizer were already constructed
to all orders in \( \theta  \). We will extend the method developed
there to the SW map for covariant derivations.

\subsubsection{Semi-classical construction}

We will first do the construction in the semi-classical limit, where
the star commutator is replaced by the Poisson bracket. As in \cite{Jurco:2000fs}
and \cite{Jurco:2001my}, we define, with the help of the Poisson
tensor \( \theta =\frac{1}{2}\theta ^{kl}\partial _{k}\wedge \partial _{l} \)
\begin{equation}
d_{\theta }=-[\cdot ,\theta ]\end{equation}
and (locally) \begin{equation}
a_{\theta }=\theta ^{ij}a_{j}\partial _{i}.\end{equation}
Note that the bracket used in the definition of \( d_{\theta } \)
is not the Schouten-Nijenhuis bracket (\ref{Schouten-Nijenhuis bracket}).
For polyvectorfields \( \pi _{1} \) and \( \pi _{2} \) it is \begin{equation}
{}[\pi _{1},\pi _{2}]=-[\pi _{2},\pi _{1}]_{S},\end{equation}
giving an extra minus sign for \( \pi _{1} \) and \( \pi _{2} \)
both even (see \ref{Quantum commutators}). Especially, we get for
\( d_{\theta } \) acting on a function \( g \)

\begin{equation}
d_{\theta }g=-[g,\theta ]=[g,\theta ]_{S}=\theta ^{kl}\partial _{l}g\partial _{k}.\end{equation}
Now a parameter \( t \) and \( t \)-dependent \( \theta _{t}=\frac{1}{2}\theta _{t}^{kl}\partial _{k}\wedge \partial _{l} \)
and \( X_{t}=X^{k}_{t}\partial _{k} \) are introduced, fulfilling

\begin{equation}
\partial _{t}\theta _{t}=f_{\theta }=-\theta _{t}f\theta _{t}\; \; \; \; \mbox {and}\; \; \; \; \partial _{t}X_{t}=-X_{t}f\theta _{t},\end{equation}
where the multiplication is ordinary matrix multiplication. Given
the Poisson tensor \( \theta  \) and the Poisson vectorfield \( X \),
the formal solutions are\begin{equation}
\theta _{t}=\theta \sum _{n=0}^{\infty }(-t\; f\theta )^{n}=\frac{1}{2}(\theta ^{kl}-t\theta ^{ki}f_{ij}\theta ^{jl}_{}+\ldots )\partial _{k}\wedge \partial _{l}\end{equation}
and

\begin{equation}
X_{t}=X\sum _{n=0}^{\infty }(-t\; f\theta )^{n}=X^{k}\partial _{k}-tX^{i}f_{ij}\theta ^{jk}\partial _{k}+\ldots .\end{equation}
\( \theta _{t} \) is still a Poisson tensor and \( X_{t} \) is still
a Poisson vectorfield, i.e.

\begin{equation}
[\theta _{t},\theta _{t}]=0\; \; \; \; \; \; \; \mbox {and}\; \; \; \; \; \; \; \; [X_{t},\theta _{t}]=0.\end{equation}
 For the proof see \ref{Commutator theta theta and theta X}.

With this we calculate\begin{equation}
\label{f theta}
f_{\theta }=\partial _{t}\theta _{t}=-\theta _{t}f\theta _{t}=-[a_{\theta },\theta ]=d_{\theta }a_{\theta }.
\end{equation}
We now get the following commutation relations

\begin{eqnarray}
{[}a_{\theta _{t}}+\partial _{t},d_{\theta _{t}}(g)] & = & d_{\theta _{t}}((a_{\theta _{t}}+\partial _{t})(g)),\label{comm rel 1} \\
{[}a_{\theta _{t}}+\partial _{t},X_{t}] & = & -d_{\theta _{t}}(X_{t}^{k}a_{k}),\label{comm rel} 
\end{eqnarray}
where \( g \) is some function which might also depend on \( t \)
(see \ref{Semi-classical commutators}).

To construct the Seiberg-Witten map for the gauge potential \( A_{X} \),
we first define

\begin{equation}
K_{t}=\sum ^{\infty }_{n=0}\frac{1}{(n+1)!}(a_{\theta _{t}}+\partial _{t})^{n}.\end{equation}
With this, the semi-classical gauge parameter reads \cite{Jurco:2000fs,Jurco:2001my}
\begin{equation}
\Lambda _{\lambda }[a]=K_{t}(\lambda )\Big {|}_{t=0}.\end{equation}
To see that this has indeed the right transformation properties under
gauge transformations, we first note that the transformation properties
of \( a_{\theta _{t}} \) and \( X^{k}_{t}a_{k} \) are\begin{equation}
\label{delta a theta}
\delta _{\lambda }a_{\theta _{t}}=\theta ^{kl}_{t}\partial _{l}\lambda \partial _{k}=d_{\theta _{t}}\lambda 
\end{equation}
and\begin{equation}
\label{delta X}
\delta _{\lambda }(X_{t}^{k}a_{k})=X_{t}^{k}\partial _{k}\lambda =[X_{t},\lambda ].
\end{equation}
Using (\ref{delta a theta},\ref{delta X}) and the commutation relations
(\ref{comm rel 1},\ref{comm rel}), a rather tedious calculation
(see \ref{Transformation properties of K}) shows that \begin{equation}
\delta _{\lambda }K_{t}(X_{t}^{k}a_{k})=X_{t}^{k}\partial _{k}K_{t}(\lambda )+d_{\theta _{t}}(K_{t}(\lambda ))K_{t}(X_{t}^{k}a_{k}).\end{equation}
Therefore, the semi-classical gauge potential is

\begin{equation}
A_{X}[a]=K_{t}(X_{t}^{k}a_{k})\Big {|}_{t=0}.\end{equation}

\subsubsection{Quantum construction}

We can now use the Kontsevich formality map to quantise the semi-classical
construction. All the semi-classical expressions can be mapped to
their counterparts in the \( \star  \)-product formalism without
loosing the properties necessary for the construction. One higher
order term will appear, fixing the transformation properties for the
quantum objects.

The star-product we will use is \begin{equation}
\star =\sum ^{\infty }_{n=0}\frac{1}{n!}U_{n}(\theta _{t},\, \ldots ,\theta _{t}).\end{equation}
 We define 

\begin{equation}
d_{\star }=-[\cdot ,\star ]_{G\: \: },\end{equation}
which for functions \( f \) and \( g \) reads

\begin{equation}
d_{\star }(g)\: f=[f\stackrel{\star }{,}g].\end{equation}
The bracket used in the definition of \( d_{\star } \) is the Gerstenhaber
bracket (\ref{Gerstenhaber bracket}). We now calculate the commutators
(\ref{comm rel 1}) and (\ref{comm rel}) in the new setting (see
\ref{Quantum commutators}). We get

\begin{eqnarray}
{[}\Phi (a_{\theta _{t}})+\partial _{t},d_{\star }(\Phi (g))] & = & d_{\star }((\Phi (a_{\theta _{t}})+\partial _{t})\Phi (f)),\\
{[}\Phi (a_{\theta _{t}})+\partial _{t},\Phi (X_{t})] & = & -d_{\star }(\Phi (X_{t}^{k}a_{k})-\Psi (a_{\theta _{t}},X_{t})).
\end{eqnarray}
The higher order term \( \Psi (a_{\theta _{t}},X_{t}) \) has appeared,
but looking at the gauge transformation properties of the quantum
objects we see that it is actually necessary. We get \begin{equation}
\delta _{\lambda }\Phi (a_{\theta _{t}})=\Phi (d_{\theta _{t}}\lambda )=d_{\star }\Phi (\lambda )\end{equation}
with (\ref{formality star derivation}) and (\ref{delta a theta})
and\begin{eqnarray}
\delta _{\lambda }(\Phi (X_{t}^{k}a_{k})-\Psi (a_{\theta },X_{t})) & = & \Phi ([X_{t},\lambda ])-\Psi (d_{\theta }\lambda ,X_{t})\\
 & = & [\Phi (X_{t}),\Phi (\lambda )]-\Psi ([\theta _{t},\lambda ],X_{t})\nonumber \\
 &  & +\Psi ([\theta _{t},X_{t}],\lambda )-\Psi (d_{\theta }\lambda ,X_{t})\nonumber \\
 & = & [\Phi (X_{t}),\Phi (\lambda )]\nonumber \\
 & = & \delta _{X_{t}}\Phi (\lambda ),\nonumber 
\end{eqnarray}
where the addition of the new term preserves the correct transformation
property. With

\begin{equation}
K_{t}^{\star }=\sum _{n=0}^{\infty }\frac{1}{(n+1)!}(\Phi (a_{\theta _{t}})+\partial _{t})^{n},\end{equation}
a calculation analogous to the semi-classical case gives \begin{eqnarray}
\delta _{\lambda }(K_{t}^{\star }(\Phi (X_{t}^{k}a_{k})-\Psi (a_{\theta _{t}},X_{t}))) & = & \delta _{X_{t}}K_{t}^{\star }(\Phi (\lambda ))\\
 &  & +d_{\star }(K_{t}^{\star }(\Phi (\lambda )))K_{t}^{\star }(\Phi (X_{t}^{k}a_{k})-\Psi (a_{\theta _{t}},X_{t})).\nonumber 
\end{eqnarray}
As in \cite{Jurco:2000fs,Jurco:2001my}, the NC gauge parameter is

\begin{equation}
\Lambda _{\lambda }[a]=K_{t}^{\star }(\Phi (\lambda ))\Big {|}_{t=0},\end{equation}
and we therefore get for the NC gauge potential

\begin{equation}
A_{X}[a]=K_{t}^{\star }(\Phi (X_{t}^{k}a_{k})-\Psi (a_{\theta _{t}},X_{t}))\Big {|}_{t=0},\end{equation}
transforming with

\begin{equation}
\delta _{\lambda }A_{X}=\delta _{X}\Lambda _{\lambda }-[\Lambda _{\lambda }\stackrel{\star }{,}A_{X}].\end{equation}

\section{Acknowledgements}

The authors want to thank B. Jurco and J. Wess for many interesting discussions.
We also want to thank the MPI and the LMU for their support.

\begin{appendix}

\section{Definitions}

\subsection{\label{Schouten-Nijenhuis bracket}The Schouten-Nijenhuis bracket}

The Schouten-Nijenhuis bracket for multivectorfields \( \pi ^{i_{1}\ldots i_{k_{s}}}_{s}\partial _{i_{1}}\wedge \ldots \wedge \partial _{i_{k_{s}}} \)
can be written as (\cite{Arnal:2000hy},IV.2.1):

\begin{equation}
{}[\pi _{1},\pi _{2}]_{S}=(-1)^{k_{1}-1}\pi _{1}\bullet \pi _{2}-(-1)^{k_{1}(k_{2}-1)}\pi _{2}\bullet \pi _{1},\end{equation}

\begin{equation}
\pi _{1}\bullet \pi _{2}=\sum ^{k_{1}}_{l=1}(-1)^{l-1}\pi _{1}^{i_{1}\ldots i_{k_{1}}}\partial _{l}\pi _{2}^{j_{1}\ldots j_{k_{2}}}\partial _{i_{1}}\wedge \ldots \wedge \widehat{{\partial _{i_{l}}}}\wedge \ldots \wedge \partial _{i_{k_{1}}}\wedge \partial _{j_{1}}\wedge \ldots \wedge \partial _{j_{k_{2}}},\end{equation}
 where the hat marks an omitted derivative.

For a function \( g \), vectorfields \( X=X^{k}\partial _{k} \)
and \( Y=Y^{k}\partial _{k} \) and a bivectorfield \( \pi =\frac{1}{2}\pi ^{kl}\partial _{k}\wedge \partial _{l} \)
we get:

\begin{eqnarray}
{}[X,g]_{S} & = & X^{k}\partial _{k}g,\\
{}[\pi ,g]_{S} & = & -\pi ^{kl}\partial _{k}g\partial _{l},\\
{}[X,\pi ]_{S} & = & \frac{1}{2}(X^{k}\partial _{k}\pi ^{ij}-\pi ^{ik}\partial _{k}X^{j}+\pi ^{jk}\partial _{k}X^{i})\partial _{i}\wedge \partial _{j},\\
{}[\pi ,\pi ]_{S} & = & \frac{1}{3}(\pi ^{kl}\partial _{l}\pi ^{ij}+\pi ^{il}\partial _{l}\pi ^{jk}+\pi ^{jl}\partial _{l}\pi ^{ki})\partial _{k}\wedge \partial _{i}\wedge \partial _{j}.
\end{eqnarray}

\subsection{\label{Gerstenhaber bracket}The Gerstenhaber bracket}

The Gerstenhaber bracket for polydifferential operators \( A_{s} \)
can be written as (\cite{Arnal:2000hy},IV.3):

\begin{equation}
{}[A_{1},A_{2}]_{G}=A_{1}\circ A_{2}-(-1)^{(|A_{1}|-1)(|A_{2}|-1)}A_{2}\circ A_{1},\end{equation}
\begin{eqnarray}
\lefteqn {(A_{1}\circ A_{2})(f_{1},\ldots \, f_{m_{1}+m_{2}-1})} &  & \\
 & = & \sum ^{m_{1}}_{j=1}(-1)^{(m_{2}-1)(j-1)}A_{1}(f_{1},\ldots \, f_{j-1},A_{2}(f_{j},\ldots ,f_{j+m_{2}-1}),f_{j+m_{2}},\ldots ,f_{m_{1}+m_{2}-1}),\nonumber 
\end{eqnarray}
where \( |A_{s}| \) is the degree of the polydifferential operator
\( A_{s} \), i.e. the number of functions it is acting on.

For functions \( g \) and \( f \), differential operators \( D_{1} \)and
\( D_{2} \) of degree one and \( P \) of degree two we get

\begin{eqnarray}
{}[D,g]_{G} & = & D(g),\nonumber \\
{}[P,g]_{G}(f) & = & P(g,f)-P(f,g),\nonumber \\
{}[D_{1},D_{2}]_{G}(g) & = & D_{1}(D_{2}(g))-D_{2}(D_{1}(g)),\nonumber \\
{}[P,D]_{G}(f,g) & = & P(D(f),g)+P(f,D(g))-D(P(f,g)).\label{gerst 2 1} 
\end{eqnarray}

\section{Calculations}

\subsection{\label{Commutator theta theta and theta X}Calculation of \protect\( [\theta _{t},\theta _{t}]\protect \)
and \protect\( [\theta _{t},X_{t}]\protect \)}

We want to show that \( \theta _{t} \) is still a Poisson tensor
and that \( X_{t} \) still commutes with \( \theta _{t} \). For
this we first define \( \theta (n)^{k}_{l}=(\theta f)^{n}=\theta ^{ki}f_{ij}\ldots \theta ^{rs}f_{sl}=f_{li}\theta ^{ij}\ldots f_{rs}\theta ^{sk}=(f\theta )^{n} \)
and \( \theta (n)^{kl}=\theta (f\theta )^{n}=\theta ^{ki}f_{ij}\ldots f_{rs}\theta ^{sl} \).
In the calculations to follow we will sometimes drop the derivatives
of the polyvectorfields and associate \( \pi ^{k_{1}\ldots k_{n}} \)
with \( \pi ^{k_{1}\ldots k_{n\: \: }}\frac{1}{n}\partial _{k_{1}}\wedge \ldots \wedge \partial _{k_{n}} \)for
simplicity. All the calculations are done locally.

We evaluate

\begin{eqnarray}
{[}\theta _{t},\theta _{t}]_{S} & = & \theta _{t}^{kl}\partial _{l}\theta ^{ij}_{t}+\mbox {c.p.}\; \mbox {in}\; (kij)\\
 & = & \sum ^{\infty }_{n,m=0}\sum ^{m}_{o=0}(-t)^{n+m}\theta (n)^{k}_{r}\theta (o)^{i}_{s}\theta (m-o)_{p}^{j}\theta ^{rl}\partial _{l}\theta ^{sp}+\mbox {c.p.}\; \mbox {in}\; (kij)\nonumber \\
 &  & +\sum ^{\infty }_{n,m=0}\sum ^{m}_{o=0}(-t)^{n+m+1}\theta (n)^{kl}\theta (o)^{is}\theta (m-o)^{pj}\partial _{l}f_{sp}+\mbox {c.p.}\; \mbox {in}\; (kij)\nonumber \\
 & = & \sum ^{\infty }_{n,m,o=0}(-t)^{n+m+o}\theta (n)^{k}_{r}\theta (o)^{i}_{s}\theta (m)_{p}^{j}\theta ^{rl}\partial _{l}\theta ^{sp}+\mbox {c.p.}\; \mbox {in}\; (kij)\nonumber \\
 &  & -\sum ^{\infty }_{n,m,o=0}(-t)^{n+m+o+1}\theta (n)^{kl}\theta (o)^{is}\theta (m)^{jp}\partial _{l}f_{sp}+\mbox {c.p.}\; \mbox {in}\; (kij).\nonumber 
\end{eqnarray}
The first part vanishes because \( \theta _{t} \) is a Poisson tensor,
i.e.\begin{equation}
\label{poisson}
{[}\theta ,\theta ]_{S}=\theta ^{kl}\partial _{l}\theta ^{ij}+\mbox {c.p.}\; \mbox {in}\; (kij)=0,
\end{equation}
the second part because of\begin{equation}
\label{cyclic f}
\partial _{k}f_{ij}+\mbox {c.p.}\; \mbox {in}\; (kij)=0.
\end{equation}
To prove that \( X_{t} \) still commutes with \( \theta _{t} \),
we first note that\begin{equation}
X_{t}=X\sum _{n=0}^{\infty }(-tf\theta )=X(1-tf\theta _{t}).\end{equation}
With this we can write

\begin{eqnarray}
{[}X_{t},\theta _{t}] & = & [X,\theta _{t}]-t[Xf\theta _{t},\theta _{t}]\label{comm X theta} \\
 & = & X^{n}\partial _{n}\theta _{t}^{kl}-\theta _{t}^{kn}\partial _{n}X^{l}+\theta _{t}^{ln}\partial _{n}X^{k}\nonumber \\
 &  & -tX^{m}f_{mi}\theta _{t}^{in}\partial _{n}\theta _{t}^{kl}+t\theta _{t}^{kn}\partial _{n}(X^{m}f_{mi}\theta _{t}^{il})-t\theta _{t}^{ln}\partial _{n}(X^{m}f_{mi}\theta _{t}^{ik})\nonumber \\
 & = & X^{n}\partial _{n}\theta _{t}^{kl}-\theta _{t}^{kn}\partial _{n}X^{l}+\theta _{t}^{ln}\partial _{n}X^{k}\nonumber \\
 &  & +t\theta _{t}^{kn}\partial _{n}X^{m}f_{mi}\theta _{t}^{il}-t\theta _{t}^{ln}\partial _{n}X^{m}f_{mi}\theta _{t}^{ik}\nonumber \\
 &  & +t\theta _{t}^{kn}X^{m}\partial _{n}f_{mi}\theta _{t}^{il}-t\theta _{t}^{ln}X^{m}\partial _{n}f_{mi}\theta _{t}^{ik}.\nonumber 
\end{eqnarray}
In the last step we used (\ref{poisson}). To go on we note that

\begin{equation}
t\theta _{t}^{kn}X^{m}\partial _{n}f_{mi}\theta _{t}^{il}-t\theta _{t}^{ln}X^{m}\partial _{n}f_{mi}\theta _{t}^{ik}=tX^{n}\theta _{t}^{km}\partial _{n}f_{mi}\theta _{t}^{il},\end{equation}
where we used (\ref{cyclic f}). Making use of the power series expansion
and the fact that \( X \) commutes with \( \theta , \) i.e. \begin{equation}
{[}X,\theta ]=X^{n}\partial _{n}\theta ^{kl}-\theta ^{kn}\partial _{n}X^{l}+\theta ^{ln}\partial _{n}X^{k}=0,\end{equation}
 we further get

\begin{eqnarray}
X^{n}\partial _{n}\theta _{t}^{kl}+tX^{n}\theta _{t}^{km}\partial _{n}f_{mi}\theta _{t}^{il} & = & \sum ^{\infty }_{r,s=0}(-t)^{r+s}\theta (r)^{k}_{i}X^{n}\partial _{n}\theta ^{ij}\theta (s)_{j}^{l}\\
 & = & \sum ^{\infty }_{r,s=0}(-t)^{r+s}\theta (r)^{k}_{i}\theta ^{in}\partial _{n}X^{j}\theta (s)_{j}^{l}\nonumber \\
 &  & -\sum ^{\infty }_{r,s=0}(-t)^{r+s}\theta (r)^{k}_{i}\theta ^{jn}\partial _{n}X^{i}\theta (s)_{j}^{l}.\nonumber 
\end{eqnarray}
Therefore (\ref{comm X theta}) reads

\begin{eqnarray}
{[}X_{t},\theta _{t}] & = & \sum ^{\infty }_{r,s=0}(-t)^{r+s}\theta (r)^{k}_{i}\theta (s)_{j}^{l}\theta ^{in}\partial _{n}X^{j}-\sum ^{\infty }_{r,s=0}(-t)^{r+s}\theta (r)^{k}_{i}\theta (s)_{j}^{l}\theta ^{jn}\partial _{n}X^{i}\\
 &  & -\theta _{t}^{kn}\partial _{n}X^{l}+\theta _{t}^{ln}\partial _{n}X^{k}+t\theta _{t}^{kn}\partial _{n}X^{m}f_{mi}\theta _{t}^{il}-t\theta _{t}^{ln}\partial _{n}X^{m}f_{mi}\theta _{t}^{ik}\nonumber \\
 & = & 0.\nonumber 
\end{eqnarray}

\subsection{\label{Transformation properties of K}The transformation properties
of \protect\( K_{t}\protect \)}

To calculate the transformation properties of \( K_{t}(X_{t}^{k}a_{k}) \),
we first evaluate

\begin{eqnarray}
\delta _{\lambda }((a_{\theta }+\partial _{t})^{n})X^{k}a_{k} & = & \sum ^{n-1}_{i=0}(a_{\theta }+\partial _{t})^{i}d_{\theta }(\lambda )(a_{\theta }+\partial _{t})^{n-1-i}X^{k}a_{k}\\
 & = & \sum ^{n-1}_{i=0}\sum ^{i}_{l=0}{i\choose l}d_{\theta }((a_{\theta }+\partial _{t})^{l}(\lambda ))(a_{\theta }+\partial _{t})^{n-1-l}X^{k}a_{k}\nonumber 
\end{eqnarray}
and

\begin{eqnarray}
 &  & (a_{\theta }+\partial _{t})^{n}\delta _{\lambda }(X^{k}a_{k})\\
 & = & (a_{\theta }+\partial _{t})^{n}X^{k}\partial _{k}\lambda\nonumber  \\
 & = & X^{k}\partial _{k}(a_{\theta }+\partial _{t})^{n}-\sum ^{n-1}_{i=0}(a_{\theta }+\partial _{t})^{i}d_{\theta }(X^{k}a_{k})(a_{\theta }+\partial _{t})^{n-1-i}\lambda\nonumber  \\
 & = & X^{k}\partial _{k}(a_{\theta }+\partial _{t})^{n}\nonumber \\
 &  & -\sum ^{n-1}_{i=0}\sum ^{n-1-i}_{j=0}{n-1-i\choose j}(-1)^{n-1-i-j}(a_{\theta }+\partial _{t})^{i+j}d_{\theta }((a_{\theta }+\partial _{t})^{n-1-i-j}(X^{k}a_{k}))(\lambda )\nonumber \\
 & = & X^{k}\partial _{k}(a_{\theta }+\partial _{t})^{n}\nonumber \\
 &  & +\sum ^{n-1}_{i=0}\sum ^{n-1-i}_{j=0}{n-1-i\choose j}(-1)^{n-1-i-j}(a_{\theta }+\partial _{t})^{i+j}d_{\theta }(\lambda )((a_{\theta }+\partial _{t})^{n-1-i-j}(X^{k}a_{k}))\nonumber \\
 & = & X^{k}\partial _{k}(a_{\theta }+\partial _{t})^{n}\nonumber \\
 &  & +\sum ^{n-1}_{i=0}\sum ^{n-1-i}_{j=0}\sum ^{i+j}_{l=0}{n-1-i\choose j}{i+j\choose l}(-1)^{n-1-i-j}d_{\theta }((a_{\theta }+\partial _{t})^{l}(\lambda ))((a_{\theta }+\partial _{t})^{n-1-l}(X^{k}a_{k})).\nonumber 
\end{eqnarray}
We go on by simplifying these expressions. Using\begin{equation}
\label{binomial}
{i\choose l}={i-1\choose l}+{i-1\choose l-1}\; \; \; \; \mbox {for}\; \; \; \; i>l
\end{equation}
we get\begin{equation}
\sum ^{n-1}_{m=l}\sum ^{m}_{i=0}{n-1-i\choose m-i}{m\choose l}(-1)^{n-1-m}=\sum ^{n-1}_{m=l}{n\choose m}{m\choose l}(-1)^{n-1-m}.\end{equation}
 Using (\ref{binomial}) again two times and then using induction
we go on to\begin{equation}
\sum ^{n-1}_{m=l}{n\choose m}{m\choose l}(-1)^{n-1-m}=\sum ^{l}_{i=0}{n-1-i\choose n-1-l},\end{equation}
 giving, after using (\ref{binomial}) again\begin{equation}
\sum ^{l}_{i=0}{n-1-i\choose n-1-l}={n\choose l}.\end{equation}
Together with\begin{equation}
\sum ^{n-1}_{i=l}{i\choose l}={n\choose l+1}\end{equation}
these formulas add up to give\begin{equation}
\sum ^{n-1}_{m=l}\sum ^{m}_{i=0}{n-1-i\choose m-i}{m\choose l}(-1)^{n-1-m}+\sum ^{n-1}_{i=l}{i\choose l}={n+1\choose l+1}\end{equation}
and therefore

\begin{equation}
\delta _{\lambda }(K_{t}(X^{k}a_{k}))=X^{k}\partial _{k}(K_{t}(\lambda ))+d_{\theta }(K_{t}(\lambda ))K_{t}(X^{k}a_{k}).\end{equation}

\subsection{Calculation of the commutators}

\subsubsection{\label{Semi-classical commutators}Semi-classical construction}

We calculate the commutator (\ref{comm rel 1}) (see also \cite{Jurco:2001my}),
dropping the t-subscripts on \( \theta _{t} \) for simplicity and
using local expressions.

\begin{eqnarray}
{[}a_{\theta },d_{\theta }(g)] & = & -\theta ^{ij}a_{j}\partial _{i}\theta ^{kl}\partial _{k}g\partial _{l}-\theta ^{ij}a_{j}\theta ^{kl}\partial _{i}\partial _{k}g\partial _{l}\\
 &  & +\theta ^{kl}\partial _{k}g\partial _{l}\theta ^{ij}a_{j}\partial _{i}+\theta ^{kl}\partial _{k}g\theta ^{ij}\partial _{l}a_{j}\partial _{i}\nonumber \\
 & = & -\theta ^{kl}\partial _{k}\theta ^{ij}a_{j}\partial _{i}g\partial _{l}-\theta ^{kl}\theta ^{ij}a_{j}\partial _{k}\partial _{i}g\partial _{l}-\theta ^{kl}\theta ^{ij}\partial _{j}a_{k}\partial _{i}g\partial _{l}\nonumber \\
 & = & +\theta ^{ij}f_{jk}\theta ^{kl}\partial _{i}g\partial _{l}-\theta ^{kl}\partial _{k}(\theta ^{ij}a_{j}\partial _{i}g)\partial _{l}\nonumber \\
 & = & -d_{\theta f\theta }g+d_{\theta }(a_{\theta }(g))\nonumber \\
 & = & -\partial _{t}(d_{\theta })g+d_{\theta }(a_{\theta }(g)).\nonumber 
\end{eqnarray}
For (\ref{comm rel}) we get

\begin{eqnarray}
{[}a_{\theta },X_{t}] & = & \theta ^{ij}a_{j}\partial _{i}X^{k}\partial _{k}-X^{k}\partial _{k}\theta ^{ij}a_{j}\partial _{i}-X^{k}\theta ^{ij}\partial _{k}a_{j}\partial _{i}\\
 & = & -\theta ^{ij}X^{k}\partial _{k}a_{j}\partial _{i}-\theta ^{ik}\partial _{k}X^{j}a_{j}\partial _{i}\nonumber \\
 & = & X^{k}f_{ki}\theta ^{ij}\partial _{j}+\theta ^{ij}\partial _{i}(X^{k}a_{k})\partial _{j}\nonumber \\
 & = & -\partial _{t}X-d_{\theta }(X^{k}a_{k}).\nonumber 
\end{eqnarray}

\subsubsection{\label{Quantum commutators}Quantum construction}

In \cite{Manchon:2000hy}, (\ref{comm phi f star},\ref{comm delta X star},\ref{comm delta X Phi g})
have already been calculated, unluckily (and implicitly) using a different
sign convention for the brackets of polyvectorfields. In \cite{Jurco:2001my},
again a different sign convention is used, coinciding with the one
in \cite{Manchon:2000hy} in the relevant cases. In order to keep
our formulas consistent with the ones used in \cite{Manchon:2000hy,Jurco:2001my},
we define our bracket on polyvectorfields \( \pi _{1} \) and \( \pi _{2} \)
as in \cite{Manchon:2000hy} to be\begin{equation}
{}[\pi _{1},\pi _{2}]=-[\pi _{2},\pi _{1}]_{S},\end{equation}
giving an extra minus sign for \( \pi _{1} \) and \( \pi _{2} \)
both even. The bracket on polydifferential operators is always the
Gerstenhaber bracket.

With these conventions and\begin{equation}
d_{\star }=-[\cdot ,\star ]\end{equation}

we rewrite the formulas (\ref{comm delta X Phi g},\ref{comm delta X delta Y},\ref{comm phi f star},\ref{comm delta X star})
so we can use them in the following

\begin{eqnarray}
{}[\Phi (X),\Phi (g)]_{G} & = & \Phi ([X,g])+\Psi ([\theta ,g],X)-\Psi ([\theta ,X],g),\label{Phi X Phi g} \\
{}[\Phi (X),\Phi (Y)]_{G} & = & d_{\star }\Psi (X,Y)\label{Phi X, Phi Y} \\
 &  & +\Phi ([X,Y])+\Psi ([\theta ,Y],X)-\Psi ([\theta ,X],Y),\nonumber \\
d_{\star }\Phi (g) & = & \Phi (d_{\theta }(g)),\\
d_{\star }\Phi (X) & = & \Phi (d_{\theta }(X)).\label{d star x} 
\end{eqnarray}
For the calculation of the commutators of the quantum objects we first
define\begin{equation}
a_{\star }=\Phi (a_{\theta _{t}})\end{equation}
and \begin{equation}
f_{\star }=\Phi (f_{\theta _{t}}).\end{equation}
With (\ref{d star x}) we get the quantum version of (\ref{f theta})\begin{equation}
f_{\star }=d_{\star }a_{\star }.\end{equation}
For functions \( f \) and \( g \) we get

\begin{equation}
\partial _{t}(f\star g)=\sum ^{\infty }_{n=0}\frac{1}{n!}\partial _{t}U_{n}(\theta _{t},\, \ldots ,\theta _{t})(f,g)=\sum ^{\infty }_{n=1}\frac{1}{(n-1)!}U_{n}(f_{\theta },\, \ldots ,\theta _{t})(f,g)=f_{\star }(f,g).\end{equation}
With these two formulas we can now calculate the quantum version of
(\ref{comm rel 1}) as in \cite{Jurco:2001my}. On two functions \( f \)
and \( g \) we have

\begin{eqnarray}
\partial _{t}(f\star g) & = & f_{\star }(f,g)\\
 & = & d_{\star }a_{\star }(f,g)\nonumber \\
 & = & -[a_{\star },\star ](f,g)\nonumber \\
 & = & -a_{\star }(f\star g)+a_{\star }(f)\star g+f\star a_{\star }(g),\nonumber 
\end{eqnarray}
where we used (\ref{gerst 2 1}) in the last step. Therefore\begin{eqnarray}
{}[a_{\star },d_{\star }(g)](f) & = & a_{\star }(d_{\star }(g)(f))-d_{\star }(g)(a_{\star }(f))\\
 & = & a_{\star }([f\stackrel{\star }{,}g])-[a_{\star }(f)\stackrel{\star }{,}g]\nonumber \\
 & = & -\partial _{t}[f\stackrel{\star }{,}g]-[a_{\star }(g)\stackrel{\star }{,}f]\nonumber \\
 & = & -\partial _{t}d_{\star }(g)(f)+d_{\star }(a_{\star }(g))(f).\nonumber 
\end{eqnarray}
For a function \( g \) which might also depend on \( t \) the quantum
version of (\ref{comm rel 1}) now reads\begin{equation}
{}[a_{\star }+\partial _{t},d_{\star }(g)]=d_{\star }(a_{\star }(g)).\end{equation}
We go on to calculate the quantum version of (\ref{comm rel}). We
first note that\begin{equation}
\partial _{t}\Phi (X_{t})=\sum ^{\infty }_{n=1}\frac{1}{(n-1)!}\partial _{t}U_{n}(X_{t},\theta _{t},\, \ldots ,\theta _{t})=\Phi (\partial _{t}X_{t})+\Psi (f_{\theta },X_{t}).\end{equation}
With this we get\begin{eqnarray}
{}[\Phi (a_{\theta }),\Phi (X_{t})] & = & d_{\star }\Psi (a_{\theta },X_{t})+\Phi ([a_{\theta },X_{t}])-\Psi ([\theta _{t}\, a_{\theta }])+\Psi ([\theta _{t},X_{t}],a_{\theta })\\
 & = & d_{\star }\Psi (a_{\theta },X_{t})+\Phi (-d_{\theta }(X_{t}^{k}a_{k}))+\Phi (-\partial _{t}X_{t})-\Psi (f_{\theta },X_{t})\nonumber \\
 & = & -d_{\star }(\Phi (X_{t}^{k}a_{k})-\Psi (a_{\theta },X_{t}))-\partial _{t}\Phi (X_{t}),\nonumber 
\end{eqnarray}
where we have used (\ref{Phi X, Phi Y}). 

\end{appendix}

\bibliographystyle{diss}
\bibliography{mainbib}

\end{document}